\documentclass[sigconf,manuscript]{acmart}
\AtBeginDocument{%
  }

\setcopyright{acmcopyright}
\copyrightyear{2025}
\acmYear{2025}
\acmDOI{10.1145/3706598.3714133}

\acmConference[CHI '25]{CHI Conference on Human Factors in Computing
Systems}{April 26-May 1, 2025}{Yokohama, Japan}
\acmBooktitle{CHI Conference on Human Factors in Computing Systems (CHI
'25), April 26-May 1, 2025, Yokohama, Japan}
\acmISBN{979-8-4007-1394-1/25/04}

\usepackage{multirow}
\usepackage{enumitem}

\definecolor{changes}{RGB}{0, 0, 0}
\begin{document}

\title{BudsID: Mobile-Ready and Expressive Finger Identification Input for Earbuds}


\author{Jiwan Kim}
\affiliation{%
  \institution{School of Electrical Engineering, KAIST}
  \city{Daejeon}
  \country{Republic of Korea}}
\email{jiwankim@kaist.ac.kr}

\author{Mingyu Han}
\affiliation{%
  \institution{Design, UNIST}
  \city{Ulsan}
  \country{Republic of Korea}}
\email{mghan@unist.ac.kr}

\author{Ian Oakley}
\affiliation{%
  \institution{School of Electrical Engineering, KAIST}
  \city{Daejeon}
  \country{Republic of Korea}}
\email{ian.r.oakley@gmail.com}


\begin{abstract}
Wireless earbuds are an appealing platform for wearable computing on-the-go. However, their small size and out-of-view location mean they support limited different inputs. We propose finger identification input on earbuds as a novel technique to resolve these problems. This technique involves associating touches by different fingers with different responses. To enable it on earbuds, we adapted prior work on smartwatches to develop a wireless earbud featuring a magnetometer that detects fields from a magnetic ring. A first study reveals participants achieve rapid, precise earbud touches with different fingers, even while mobile (time: 0.98s, errors: 5.6\%). Furthermore, touching fingers can be accurately classified (96.9\%). A second study shows strong performance with a more expressive technique involving multi-finger double-taps (inter-touch time: 0.39s, errors: 2.8\%) while maintaining high accuracy (94.7\%). We close by exploring and evaluating the design of earbud finger identification applications and demonstrating the feasibility of our system on low-resource devices.
\end{abstract}



\begin{teaserfigure}
  \includegraphics[width=\textwidth]{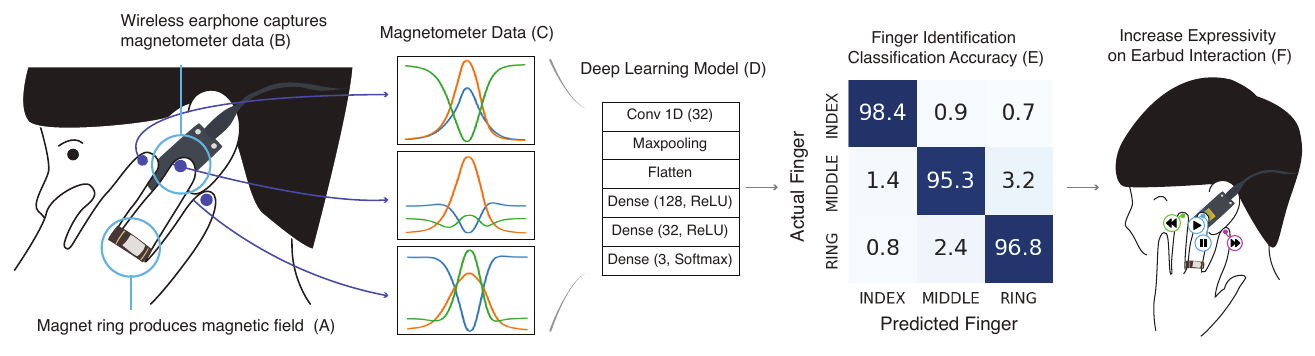}
  \caption{Overview of BudsID. A magnet ring worn on a user's middle finger produces a magnetic field (A). An earbud senses this magnetic field using an integrated magnetometer when it is tapped by different fingers (among index, middle and ring) (B). Captured magnetometer data (C) are passed to a deep learning model (D) to identify which finger performed the touch. We achieved a mean of 96.9\% accuracy in identifying single touching fingers in individual models (E). We suggest our system can increase expressivity on earbud interaction by assigning different functions to different fingers (F).}
  \Description{In Figure 1, sub-images labeled (A) and (B) show wireless earphones capturing magnetometer data, with a magnetic ring on a user's middle finger producing a magnetic field. sub-image (C) illustrates the magnetometer data plots that occur when index, middle, and ring finger are used to tap. Notably, in sub-image (C), each finger tap exhibits distinctly different patterns. (D) depicts the deep learning model we used for classification while (E) shows the confusion matrices we achieved for individual models. specifically, accuracies for index, middle and ring finger are 98.4\%, 95.3\% and 96.8\% (mean 96.9\%). The largest proportion of misclassifications are between middle and ring ringer (3.2\% and 2.4\%). Lastly, (F) illustrates a man interacting with earbuds by assigning different functions such as prev/play/next for music control to different fingers. }
  \label{fig:teaser}
\end{teaserfigure}


\maketitle

\section{Introduction}
Wireless earbuds are a popular and useful device category. Their success is predicated on the high comfort, convenience, and effectiveness of the core service they provide---playing and capturing audio while users are on the go. Numerous researchers have suggested that their prevalence, and people's fundamental willingness to wear them while out and about, make them ideal devices in which to embed other computational functionality. For example, research has demonstrated how earbuds can be used for unobtrusive health tracking and monitoring~\cite{poh2009heartphones, athavipach2019wearable, rahman2022breathebuddy, martin2017ear}, how they can support biometric authentication~\cite{wang2021eardynamic,curran2016passthoughts,gao2019earecho}, how they can control functions on a smartphone~\cite{xu2020earbuddy, kikuchi2017eartouch}, and more widely, how they can serve as general-purpose interfaces to interact with external Internet-of-Things (IoT) devices in the immediate environment~\cite{katayama2019situation}. We believe that integrating such functionality into the next generation of earbuds has the potential to transform them into general-purpose smart devices capable of supporting mobile users in a wide range of daily tasks. 

While the appeal of such scenarios is clear, the diminutive size, lack of graphical displays, out-of-sight wearing location, and frequent dynamic use scenarios (e.g., while walking) inherent to operating earbuds means they face particular challenges as traditional touch input devices. They can physically fit only a very small number of touch areas or buttons (typically one to three)~\cite{BlindSight} and, additionally, can detect swipes in the four cardinal directions~\cite{xu2020earbuddy,rateau2022leveraging}, an input modality generally considered robust to the disturbances typical during mobile use~\cite{dswime18singh, gestural02prhonen}. This limited input repertoire typically constrains the use of earbuds to control a small number of key device parameters, such as changing device volume. To extend the scenarios in which these devices can be used, researchers have begun to explore novel input modalities designed specifically for use with earbuds. For instance, EarTouch~\cite{kikuchi2017eartouch} uses optical sensors to detect ear deformation when users gently pull their ears to control applications like music players and navigation systems. In a similar context, on-face and near-ear~\cite{xu2020earbuddy} as well as in-air~\cite{rateau2022leveraging} input methods for earbuds have been proposed. While these methods aim to provide eyes-free and socially acceptable ways of interacting with devices, we note they have not yet been assessed in genuinely mobile scenarios, such as while users are walking. As such, we argue that further design explorations and systematic evaluations of earbud interaction techniques will be critical to the future development of earbud device form factor. 

One technique that holds particular promise in this area is finger identification---differentiating input based on which of a user's digits initiates a device touch. This input paradigm has been proposed for a wide range of device platforms---including computer track pads~\cite{masson2017whichfingers}, smartphones~\cite{le2019investigating} and smartwatches~\cite{kim2022sonarid,gil2017tritap,geunwoo2020magtouch}---to achieve a variety of designs. For example, the use of two different fingers during device touches can be used to trigger a primary or secondary action on a particular target, a process akin to using the left and right buttons on a mouse. In a similar vein, such systems can be used to improve text entry performance on small screens by associating each character with a specific combination of on-screen key and tapping finger, thus reducing the number of targets required to unambiguously specify all characters and enabling the use of correspondingly larger targets~\cite{gupta2016dualkey}. Alternatively, the approach can be used to support multi-tasking through the simple expedient of routing all input made by a given finger to one particular application~\cite{gupta2016porous, metzger2004freedigiter}. These designs showcase the elegance, ease, and versatility of finger identification input and highlight how it is particularly suitable for small wearable devices. 

However, despite the inherent suitability of the basic technique, we are not aware of any prior work that has investigated finger identification systems on earbuds. This paper addresses this omission and presents the first design, implementation, and evaluation of finger identification input on earbuds. To achieve this, we adopted a best-in-class approach to finger identification proposed in prior work on commercial smartwatches~\cite{geunwoo2020magtouch}. Figure~\ref{fig:teaser} illustrates an overview of our system. It relies on a magnetometer in the device and requires users to wear a magnetic ring on their middle finger. We note instrumentation of the finger is a common approach used in numerous prior studies of finger identification input~\cite{gupta2016dualkey, masson2017whichfingers, gupta2016porous}.
Additionally, \textcolor{changes}{although research prototypes demonstrate the feasibility of integrating magnetometers in earbuds~\cite{montanari2024omnibudssensoryearableplatform}, there are currently no off-the-shelf magnetometer-equipped earbuds. Accordingly,} we developed an open-source wireless earbud prototype for this work. The sensing principle is simple: as different fingers move to touch the earbuds, they cause disturbances in the sensed magnetic field that are sufficiently different to reliably identify the touching finger. 

We evaluate the potential of finger identification on earbuds using this system through two studies, each of which is designed to tackle a fundamental issue---firstly, the impact of mobility on performance and, subsequently, increasing the expressivity of the technique. We achieve this via an initial study (N=24) that explores performance with finger identification input in which users tap buds on both ears with their index, middle, and ring fingers while sitting, standing, and walking a circuit. This characterizes basic performance with the technique on the novel out-of-sight location of the ear and also documents the impacts of both pose and mobility. The results (mean task time of 0.98 seconds and errors of 5.6\%) suggest that finger identification is appropriate for earbuds and, additionally, that is robust to the disturbances inherent in walking. This is a novel and highly positive result. Additionally, we explore the design of classifiers to disambiguate the touching finger, ultimately attaining peak mean performance of 96.9\% in individual models, and discuss the factors contributing to this best-in-class performance. We extend these findings in our second study (N=16) which implements a novel \textit{multi-finger double-tap} technique that increases the expressivity of finger identification input by allowing users to combine a pair of rapid touches performed by different fingers to achieve an additional nine distinct input actions. We find users can achieve fast input (inter-touch time: 0.4s) and low errors (2.8\%) and that, with a carefully designed classifier, all nine inputs remain readily distinguishable (peak mean accuracy of 94.7\% in individual models). By enabling a richer set of inputs, this study highlights the potential of finger identification input to support sophisticated input and interaction scenarios on earbuds. Building on these results, we close the paper by presenting and evaluating a set of application designs and input techniques that instantiate and realize these scenarios and showcase how finger identification input could be integrated into future earbud devices. Our contributions include:

\begin{itemize} [labelindent=\parindent,itemindent=0pt,leftmargin=*]

\item An open-source, AI-enabled wireless earbud sensing platform built using off-the-shelf microprocessors and components and casings that can be fabricated on low-end 3D printers. This enables other researchers to replicate and extend our work. 

\item User performance data from the first studies of finger identification on earbuds, a novel scenario involving both hands and out-of-sight input. \textcolor{changes}{This includes observations relating to comfort and usability of different earbud finger touches and rigorous performance data during the previously unstudied scenario of mobile use, which is particularly relevant for earbuds. This comprehensive data documents the fundamental viability of finger identification input on earbuds.}

\item \textcolor{changes}{Design and validation of a novel compound multi-finger double-tap interaction technique that increases the expressivity of finger identification. This technique can be used or further developed by other designers and researchers.} 

\item Technical data on the design and performance of features and classifiers for disambiguating the fingers touching an earbud based on the magnetic signatures evoked by a worn ring magnet. \textcolor{changes}{In addition, documentation regarding the relative impact of different sources of environmental noise experienced during magnetic sensing.} These insights can inform the design of future sensing systems to achieve finger identification input on earbuds.
\end{itemize} 
\section{Related Work}
\subsection{Interaction on Earbuds}
Wireless earbuds have gained popularity for their audio capabilities, but recent research explores their potential to support broader computational functionality. For example, work has investigated earbuds' capacity for health tracking and monitoring~\cite{poh2009heartphones, rahman2022breathebuddy}, biometric authentication~\cite{choi2023earppg, wang2021eardynamic}, smartphone control~\cite{xu2020earbuddy,kikuchi2017eartouch}, and motion tracking~\cite{gong2021robust}. These projects envision earbuds as versatile smart devices supporting users in various daily tasks. To enable this richer role, research has also proposed a range of novel input and interaction techniques for earbuds. Earput~\cite{lissermann2014earput}, for example, explored how accessories worn behind the ear, such as headsets, clips, or glasses, could serve as innovative interfaces that allow interaction on or with the human ear. However, these systems supported relatively inexpressive touch input composed of a small number of unique touch areas and cardinal swipes. This limited their potential. More recent systems show greater promise. For instance, EarBuddy~\cite{xu2020earbuddy} detects tapping and sliding gestures near the face and ears using the microphone in a commercial wireless earbud, a richer, eyes-free, and potentially socially acceptable~\cite{Lee2018SocialAcc} input method. The work in this paper builds on this prior work and further explores interaction on earbuds by examining both user and system performance during finger identification input. 

\subsection{Finger Identification}
Finger identification, a simple, effective interaction technique that has been widely applied in smartphones~\cite{le2019investigating}, smartwatches~\cite{geunwoo2020magtouch, gil2017tritap, kim2022sonarid, watchSense17Sridhar}, and tabletop computers ~\cite{au2010multitouch, benko2009enhancing}, aims to increase interaction expressivity~\cite{roy2015glass+} by differentiating input based on the tapping finger. It supports a wide range of interface designs~\cite{goguey2016performance, masson2017whichfingers}. For instance, WhichFingers~\cite{masson2017whichfingers} proposed identifying fingers during single touches or two simultaneous touches on keyboards and touchpads for tasks such as pointing, scrolling, and window management. Also, identifying consecutive tapping fingers allows for primary and secondary actions triggered by different fingers, improving text entry~\cite{gupta2016dualkey} and multi-tasking capabilities~\cite{gupta2016porous}. Finger identification is reported to be especially valuable for small wearable devices, such as smartwatches, as their small size limits the expressiveness of conventional input techniques~\cite{siek2005fat}. Our paper extends the application of finger identification to a novel wearable device form factor: wireless earbuds. By doing so, we plan to shed light on the effectiveness with which users can perform different finger taps on earbuds, a potentially more challenging task involving input on an out-of-sight device. Additionally, we further extend prior work by considering two critical design issues of particular relevance to the earbud form factor. First, we examine performance in typical mobility conditions, reflecting the fact that earbuds are often used while on the go. Secondly, we design and evaluate a rapid multi-finger double-tap technique, which is specifically designed to increase the expressivity of the finger identification input on small, screen-less devices. In these ways, we explore the viability of the finger identification technique for enhancing the expressiveness of input on earbuds.

Finger identification has been implemented via a wide range of approaches. Many involve instrumenting individual fingers with active devices, such as a distance sensors~\cite{Multitouch1, Multitouch2} or vibrating actuators~\cite{masson2017whichfingers}, or, more practically, passive devices such as magnets~\cite{geunwoo2020magtouch}. The primary goals of this body of work are understanding human performance with the technique, expanding on the design space it enables, and gaining insights into the cues that can be leveraged to reliably identify touching fingers. Research has also moved beyond these explorations to study finger identification approaches that do not require finger-worn devices. Such approaches include examining variations in the touch patterns created by each finger on a capacitive sensor grid~\cite{gil2017tritap, le2019investigating} or the sonar scene around a device~\cite{kim2022sonarid}. While these more technical studies target laudably practical goals, they typically result in relatively reduced accuracy~\cite{gil2017tritap} or require complex computational processing that makes them unsuitable for deployment on small wearable devices~\cite{le2019investigating, kim2022sonarid}. In this initial study of finger identification input on earbuds, our goals are more closely aligned with work that seeks to understand human performance with the technique and document its potential. As such, we opted to enable finger identification input by deploying a previously proposed sensing approach based on augmenting the finger with a simple, practical, passive magnetic ring~\cite{geunwoo2020magtouch}. 

\subsection{Interaction Using Magnetic Field Sensing}
The use of magnetic field sensing to enable novel forms of input has been widely explored on both mobile~\cite{cagri2015expanding, kadomura2015magnail} and wearable platforms~\cite{geunwoo2020magtouch}. One key advantage of the approach is its ability to track passive objects. Although the item or body part being tracked does need to be instrumented, no electronics or power are required: embedding or wearing a simple, and usually small and light, magnet is sufficient. Moreover, functionality to sense magnetic fields is mature and widely built into the existing Inertial Measurement Unit (IMU) modules commonly integrated into consumer devices such as smartphones and smartwatches. In addition, sensing changes to magnetic fields represents a rich sensing modality that can support a wide range of input techniques. For example, magnetic sensing can enrich touch interaction~\cite{kadomura2015magnail, liang2013gaussbits} through magnets mounted on a user's nails, and such a setup is reported to be sufficiently high-fidelity to support both rich gestural input~\cite{harrison2009abracadabra, mcintosh2019magnetips} and 3D finger tracking~\cite{chen2013utrack, chen2016finexus}. While much prior work has focused on placing the magnet on the fingertip, the potential of magnetic rings has also been explored by several authors~\cite{Ashbrook11Nenya}. In work closely related to the current study, Park et al.~\cite{geunwoo2020magtouch} investigated the use of finger-worn magnet rings to enhance touch interaction by identifying the specific fingers involved in touches to a smartwatch. This study builds upon this design, exploring the viability of finger identification on wireless earbuds using a magnetometer in the earbud and a magnet ring worn on the finger.

\section{System}
\label{section:system}
In this section, we describe the hardware, software, and classifiers used in this work. All materials to produce and replicate our work (e.g., 3D models, code, scripts, binaries) are available for download~\footnote{https://github.com/witlab-kaist/BudsID}.

\begin{figure}[t!]
\centering
   \includegraphics[width=\textwidth]{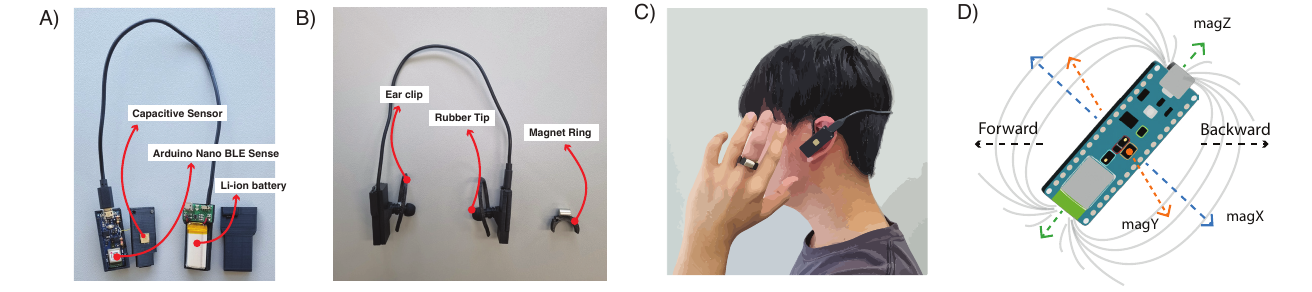}
   \hfil
\caption{BudsID prototype and sensing direction: plan views of disassembled (A) and assembled (B) earbuds. When a user wears BudsID (C), a magnetometer in an earbud-mounted microcontroller measures tri-axial variations in the surrounding magnetic field (D).}
\Description{In figure 2, Photos labeled (A) depict the disassembled hardware of BudsID earbuds, which include a capacitive sensor, Arduino Nano BLE Sense, and a Li-ion battery. Photo (B) shows an assembled earbud and a magnet ring. (C) depicts a man wearing BudsID, and (D) describes how a magnetometer in an earbud-mounted microcontroller measures tri-axial variations in the surrounding magnetic field}
\label{fig:hardware}
\end{figure}

\subsection{Hardware}
\textcolor{changes}{Magnetometers are not yet integrated into commercial earbuds due to challenges in orientation detection due to the presence of built-in magnets in speaker coils and docking connectors. However, ongoing research aims to overcome these limitations through techniques such as automatic calibration procedures capable of accommodating complex fixed and variable magnetic offsets~\cite{Ferlini21} with the goal of supporting key applications such as navigation~\cite{PilotEar21Ferlini, 9826109}. Progress towards this goal is evidenced by the recent introduction of OmniBuds~\cite{montanari2024omnibudssensoryearableplatform}, an earbud prototype with an integrated magnetometer supporting inertial tracking and device orientation sensing. Based on their utility for core wearable applications such as navigation and exercise monitoring, we expect to see magnetometers appear in commercial devices in the near future. Accordingly}, to pursue our research, we designed and developed a wireless prototype built around the Arduino Nano BLE Sense\footnote{https://docs.arduino.cc/hardware/nano-33-ble-sense}, a small (18 mm by 45 mm) AI and BLE-enabled development board featuring the LSM9DS1\footnote{https://www.st.com/en/mems-and-sensors/lsm9ds1.html}, a 9-DOF IMU with an integrated magnetometer, gyroscope and accelerometer. To create our prototype, we first configured the magnetometer to operate at 80 Hz and the accelerometer and gyroscope to operate at 104 Hz (the default); all other settings were also left at their defaults. We then designed a minimal enclosure for this board that incorporated a single 7mm square capacitive sensor\footnote{https://playground.arduino.cc/Main/CapacitiveSensor/} in the form of a copper plate in the center of its top surface. We designed an around-ear clip for this enclosure that achieved a comfortable fit and orientated the device sloping down towards the mouth (a typical design for earbuds). We used commercially available interchangeable rubber earbud tips to secure the device in the external ear canal and also ensure that the entrance to the ear canal was also positioned in the center of the device, directly underneath the capacitive sensor. In this way, the ear canal entrance served as a bodily landmark marking the location of our touch sensor. 

To provide power for this device, we designed a second earbud unit for the opposite ear featuring a 3.7V 350mAh Li-ion battery, a bespoke battery charger, and a 5V voltage booster\footnote{https://www.adafruit.com/product/4654}. We connected this unit to the Arduino Nano BLE Sense via its USB port and arranged the cabling such that it would fit comfortably behind the wearer's head. Finally, we constructed a second prototype with the sensing and battery earbud units both designed for the opposite ear. In this way, one prototype featured the sensors on the left ear and power on the right, and the other prototype, vice versa. The sensor earbud weighed 8.9 grams and the battery earbud, including the cable, weighed 25.3 grams. These weights do not substantially encumber the ear. Additionally, we produced a simple ring containing a 10mm long by 8mm diameter N35 neodymium magnet with a surface Gauss of approximately 4200, \textcolor{changes}{substantially in excess of the surface magnetic field strengths of commercial earbuds (mean 139.0, SD 88.1)~\cite{MAKINISTIAN2022113907}.} We selected the size, strength, and orientation of the magnet following recommendations from prior work~\cite{geunwoo2020magtouch}. An elastic band secured the ring to ensure a firm fit regardless of a user's hand and finger size. Figure~\ref{fig:hardware} shows the earbuds and ring hardware and how the prototype is worn. All parts for our prototype (save the rubber earbud tips, which are commonly available) were 3D printed in skin-safe PLA and can be assembled via a combination of click fit and standard M1 and M2 nuts and bolts.

\begin{figure*}[t!]
\centering
   \includegraphics[width=\textwidth]{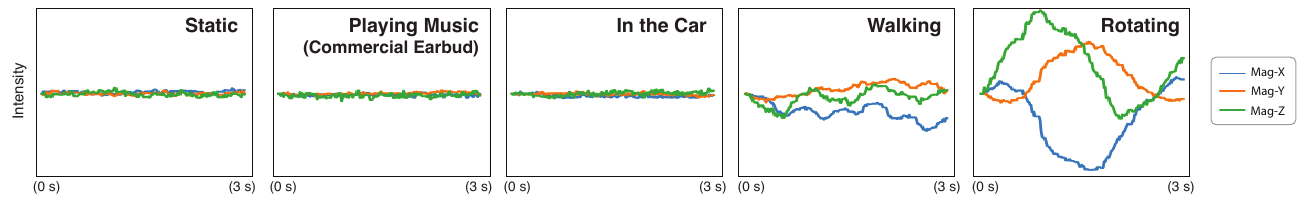}
   \hfil
\caption{\textcolor{changes}{Sample magnetometer data captured in the five different potentially noisy environments. 1. Static: static sitting posture, 2. Playing Music: wearing BudsID mounted on a commercial earbud playing music at a default (mid-range, comfortable) volume, 3. In the car: seated in a car, a potential source of magnetic interference, 4. Walking: walking in a straight line, and 5. Rotating: rotating clockwise in place.}}
\Description{ Figure 3 illustrates sample magnetometer data captured in the five different potentially noisy environments. 1. Static: static sitting posture, 2. Playing Music: wearing BudsID mounted on a commercial earbud playing music at a default (mid-range, comfortable) volume, 3. In the car: seated in a car, a potential source of magnetic interference, 4. Walking: walking in a straight line, and 5. Rotating: rotating clockwise in place.}
\label{fig:noisy}
\end{figure*}

\subsection{Software}
We developed software for our earbuds device that sampled data from our capacitive and IMU sensor at 80 Hz and transmitted them to a host PC over BLE. Due to latencies inherent in BLE, data transmission rates were both somewhat variable and somewhat reduced. Specifically, the PC polled for updated sensor data at 60Hz, while, in practice, data was received at a mean rate of between 34Hz (SD: 2.7) when streaming all IMU data (as in study 1) and 41Hz (SD: 3.02) while streaming only magnetometer data (in study 2). To maintain data continuity, the PC automatically filled in missing data gaps using a forward-fill approach.

\subsection{Robustness to Environmental Noise}
\label{section:noise}
\textcolor{changes}{
To assess the robustness of magnetometer data to external noise, we tested the system in five environments: static (baseline), sitting in a car (steel and electronics present), playing music on commercial earbuds (potential interference from the magnets in the speaker coil), walking, and rotating in-place (two representative examples of movement-induced magnetic field variability). Figure~\ref{fig:noisy} illustrates the results. It shows that static, car, and music playback scenarios generated negligible magnetic field distortions. In contrast, the movement-based scenarios of walking and rotating caused the largest disruptions, highlighting the need to validate our system in different mobility conditions to assess its real-world viability. }

\subsection{Sensing Principles}
To better understand sensing principles underlying our system, figure ~\ref{fig:hardware}-D illustrates the orientation of the three perpendicular magnetic field sensing axes of the magnetometer: these are tilted, as our prototype is aligned diagonally on the ear, pointing towards the mouth (see figure ~\ref{fig:hardware}-C). Consequently, movements of a finger-mounted magnet along the forward axis of the head (e.g., \textit{forward} towards the face, or \textit{backward} away from it) lead to variations, including inversions in polarity, in data sensed by both the X and Z-axes of the magnetometer. In contrast, data sensed on the Y-axis simply varies in magnitude but not polarity. As the position of a finger-mounted magnet on the middle finger will differ, with respect to its forward-backward location relative to the ear, when tapping the buds with different fingers, we are able to use the resulting differences in the magnetic field readings to develop a finger identification classification system.

\begin{figure*}[t!]
\centering
   \includegraphics[width=\textwidth]{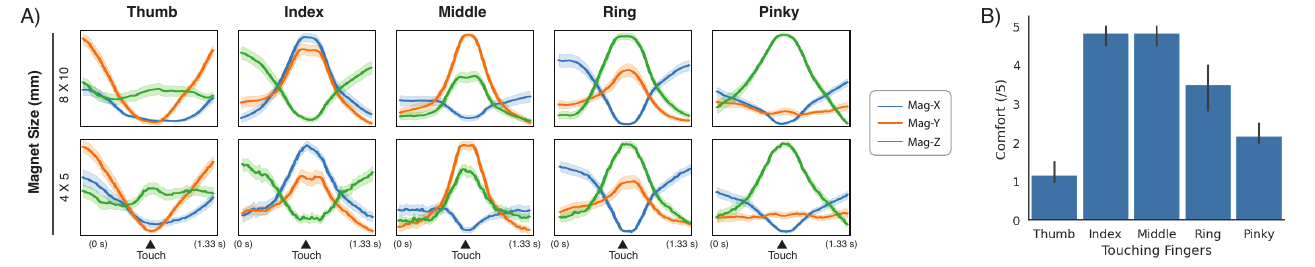}
   \hfil
\caption{\textcolor{changes}{A) The mean normalized magnetometer data captured during single taps by five different fingers with two different sizes of magnets in the pilot study (N=6) and B) mean comfort ratings for taps by each finger}}
\Description{
Figure 4 shows the study result of pilot study. A) shows the mean normalized magnetometer data captured during single taps by five different fingers with two different sizes of magnets in the pilot study, and B) shows mean comfort ratings for taps by each finger.
}
\label{fig:5fingers}
\end{figure*}

\subsection{Data Preprocessing}
Data preprocessing in our system was simple. For each input event, we first window data around its temporal center by a specific number of samples (e.g., 60, 80, or 100, a variable subsequently referred to as \textit{nSamples}). We then conduct min-max normalization on each axis separately and implement a simple polarity compensation algorithm to deal with inevitable variations in the earbud used (i.e., flipping the y-axes data to deal with the flipped sensor in left and right buds) and for inversions of the ring on the finger (i.e., flipping data from all three axes). We then followed this procedure to extract all touches recorded in our first study (see Section~\ref{section:study1}). 

\subsection{Finger Identification Viability}
\label{section:5fingers}
\textcolor{changes}{To assess the viability of the sensed magnetic field data to support finger identification, we conducted a pilot study with six right-handed participants (3 males, all right-handed). Participants wore a magnet ring on their dominant hand's middle finger and tapped the earbud using all five of their fingers (thumb, index, middle, ring, and little) with two different magnet sizes ($10\times8$ mm and $5\times4$ mm, with a measured surface Gauss of 4200 and 3600). Each participant completed two blocks, one for each magnet size, by making ten taps per finger in a random order. After the study, participants reported their perceived comfort for taps with each finger. Figure~\ref{fig:5fingers}-A displays the average captured magnetometer data recorded during single taps with all five fingers (with \textit{nSamples} set to 80). Visual examination of these images indicates that the magnetometer data generated by these different events is likely distinct enough to enable accurate finger classification, even with the smaller magnets. Comfort ratings, however, revealed strong preferences. Participants found index and middle finger taps (4.83/5) to be very comfortable and ring finger taps to be moderately comfortable (3.5/5). On the other hand, little finger taps (2.17/5) and, particularly, thumb taps (1.17/5) were rated as moderately to very uncomfortable. These findings align with prior studies highlighting the discomfort of using thumbs~\cite{gil2017tritap} or little fingers~\cite{FingerText21Lee} for wearable input and suggest these two digits may best be avoided. 
}

\subsection{Classifier}
Based on these insights, we explored the features and classifiers that can reliably identify the fingers involved in earbud touches. Specifically, we focused on two types of classifier, each of which could operate on variably sized windows of sensor data. The first approach used deep-learning techniques and the raw sensor data as features, while the second used more traditional machine-learning approaches and a set of statistical features summarizing the sensor data. The deep-learning system was intentionally lightweight: a 1D CNN classifier consisting of a single 1D convolutional layer (the number of filters: 32, kernel size: 3, stride: 2, ReLU activation) followed by max-pooling layers (pool size: 2). After flattening the output from the convolution block, we fed them into two fully connected dense layers with, respectively, 128 and 32 units. Finally, a fully connected layer was used for the multi-class classification problem of distinguishing between touching fingers. Following closely related prior work, the machine learning classifier was a support vector machine (SVM) with the radial basis function (RBF) kernel~\cite{geunwoo2020magtouch}. From each window of sensor data, we extracted six statistical features (mean, standard deviation, three quartiles (25\%, 50\%, 75\%), and magnitude) from each of the three axes of magnetometer data. Consequently, a total of 18 features were used to predict the fingers involved in earbud touches. To optimize and evaluate this classifier, we used a grid search method with cross-validation.
\section{Study 1: Finger Classification on Single Tap}
\label{section:study1}
We conducted a study to gather IMU and touch data from our system. The goals were, firstly, to assess user performance in the \textcolor{changes}{previously unstudied} task of touching earbuds with different fingers in different poses and mobility scenarios and, secondly, to support the development and evaluation of classifiers for finger identification on earbuds. This study was approved by the local Institutional Review Board (IRB).

\subsection{Participants}
We recruited 24 participants (12 male, mean age of 21.21 (SD 2.86)) from the local university through flyers. Twenty-three participants were right-handed. The participants had an average hand length of 17.83cm (SD 0.92) and a palm width of 8cm (SD 0.46). The mean length of their index, middle, and ring fingers was, respectively, 7.20cm (SD 0.37), 7.83cm (SD 0.44), and 7.08cm (SD 0.41). The study took approximately one hour to complete, and participants were compensated with approximately 15 USD in local currency for their participation.

\subsection{Design}
The study featured three independent variables: hand (right/left), finger (index/middle/ring), and posture (sit/stand/walk). We examined both hands as earbuds are typically used on both ears and thus with both hands; as such, handedness may impact performance. \textcolor{changes}{We selected the index, middle, and ring fingers based on their use in closely related prior work~\cite{geunwoo2020magtouch}. We excluded the thumb and little finger based on the poor comfort ratings we recorded for these fingers during our pilot study (see Section~\ref{section:5fingers}).} Finally, we examined different mobility conditions, as earbuds are frequently used when out and about (e.g., when traveling or exercising). \textcolor{changes}{In addition, studying different mobility conditions allows us to examine the robustness of classifiers we develop in light of variations to the sensed ambient magnetic fields due to participants turning around or otherwise changing their orientation (see Section~\ref{section:noise})}. In the \textit{sit} posture, participants were seated in a chair without armrests. In the \textit{stand} posture, participants stood still during individual study trials but were asked to rotate approximately 90 degrees clockwise between each trial. This captured performance in different orientations. Finally, during the \textit{walk} posture, participants continuously walked, at a self-defined comfortable pace, around a 16-meter (4 meters per side) square route marked out in the study room by a set of desks. This condition inherently included both changes in user orientation and also disturbances due to gait; it closely follows best practices from prior work examining input while mobile~\cite{Walking12Lim, Dobbelstein17, FingerText21Lee, Namnakani23}. 

Conditions in the study were organized as follows. We first balanced the hand variable, with half of the participants completing all left trials before all right trials and half vice versa. Within each hand order group, we fully balanced the posture variable, with two participants completing each of the six possible orders. Each hand-posture condition consisted of 60 trials, 20 of which involved each finger. These trials were delivered in a random order, and each trial involved a single-tap to the earbud's capacitive sensor. In this way, we captured data from 8640 finger taps: 24 participants by 2 hand sessions by 3 postures by 3 fingers by 20 trials.  

\subsection{Procedure}
The experiment was conducted in an empty classroom. The study commenced with participants reading instructions, signing an informed consent form, and having their hand size measured. Subsequently, they donned the earbuds and magnetic ring (on the appropriate ear and finger) and proceeded to practice tapping the earbuds while seated. They were given approximately 3 minutes for this task. They were encouraged to ask any questions during this period. Afterward, they adopted the appropriate posture (among sit, stand, and walk), and the study started. Each trial in the study began with a verbal request from the system (delivered on a speaker located in the room) to tap with any finger on the earbud touch sensor. After a touch was registered, a second verbal request asked participants to touch the buds with a specific finger by simply naming that digit (e.g., "Index"). Making a further touch to the capacitive sensor completed the trial; in cases where an initial touch failed to make contact with the sensor (e.g., the participant missed), they were instructed to make additional attempts. After a trial was completed, the next trial immediately began. Participants were asked to lower their hands between touches to reduce fatigue. In addition, following prior work in this area, videos of study tasks were recorded, but we did not independently verify tapping behavior~\cite{roy2015glass+} and instead instructed participants to verbally self-report~\cite{gil2017tritap, kim2022sonarid} if they tapped with the wrong finger. After the end of each hand-mobility condition, participants reported their workload on the NASA Task Load Index (TLX), a process that also enforced a short break. At the end of the study, participants provided demographic information and feedback on the usability, comfort, and speed of the input tasks on a 5-point Likert scale (e.g., 1 (very unsatisfied) to 5 (very satisfied)). They also discussed any uncertainty or confusion they experienced related to the task of tapping with the three different fingers. 

\begin{table*}[t!]
\centering
\small
\caption{Error rate (\%) and trial completion time (seconds) organized to show data for each of the three variables (posture, finger, and hand) in the user study. For each metric and each level of each variable, we report both mean ($\mu$) and standard deviation ($\sigma$)}
\label{table_touchbio}
\Description{Table 1 presents data from a user study organized to examine the impact of three variables---posture, finger, and hand---on error rates (\%) and trial completion times (seconds). The table is structured as follows, Variables: The top row of the table lists the three variables: "Posture," "Finger," and "Hand." Variable Levels: Under each variable, the table lists the different levels or conditions that were tested. For "Posture," the levels are "Sit," "Stand," and "Walk." For "Finger," the levels are "Index," "Middle," and "Ring." For "Hand," the levels are "Left" and "Right." Metrics: The table reports two metrics: Trial time (s) and Error rate (\%). For each metric, it reports the mean (µ) and standard deviation (σ) for each level of the variables. In summary, the table shows that trial completion times were shortest for the "Sit" posture (0.95 seconds) and error rates were lowest for the "Left" hand (4.88\%), while the "Stand" posture had the longest completion time (1.01 seconds) and the "Middle" finger exhibited the highest error rate (6.01\%).}

\begin{tabular}{l | cc|cc|cc | cc|cc|cc| cc|cc }
 & 
  \multicolumn{6}{c|}{\textbf{Posture}} & \multicolumn{6}{c|}{\textbf{Finger}} & \multicolumn{4}{c}{\textbf{Hand}}\\

 & 
 \multicolumn{2}{c|}{\textbf{Sit}}   & \multicolumn{2}{c|}{\textbf{Stand}}  & \multicolumn{2}{c|}{\textbf{Walk}} & 
 \multicolumn{2}{c|}{\textbf{Index}} & \multicolumn{2}{c|}{\textbf{Middle}} & \multicolumn{2}{c|}{\textbf{Ring}} & 
 \multicolumn{2}{c|}{\textbf{Left}}  & \multicolumn{2}{c}{\textbf{Right}}\\ 

& $\mu$ & $\sigma$ & $\mu$ & $\sigma$ & $\mu$ & $\sigma$ & $\mu$ & $\sigma$ & $\mu$ & $\sigma$ & $\mu$ & $\sigma$ & $\mu$ & $\sigma$ & $\mu$ & $\sigma$\\
 
\hline
Trial time (s)  & 0.95 & 0.18 & 1.01 & 0.2 & 0.99  & 0.22 & 1.02 & 0.2 & 0.96 & 0.21 & 0.97 & 0.18 & 1.01 & 0.21 & 0.96 & 0.19\\
Error rate (\%)  & 6.04 & 3.07 & 5.14 & 3.16 & 5.62 & 3.53 & 5.73 & 2.36 & 6.01 & 3.67 & 5.07 & 3.42 & 4.88 & 2.61 & 6.32 & 3.59	\\
\end{tabular}
\end{table*}

\subsection{User Performance Results}
\label{section:study1result}
Of the 8640 trials in the study, participants self-reported tapping the earbud incorrectly (e.g., with the wrong finger) 46 times (0.53\%), and there were 26 trials (0.3\%) subject to technical failures (mainly with the BLE link). We excluded these trials from subsequent analysis due to their low occurrence rates. Consequently, we retained 8568 trials for further analysis. With this data, we first sought to understand how effectively users can tap with different fingers on earbuds. This is important as we examine a previously unstudied version of this task that requires precise touches to a small (7mm square) out-of-sight target in conditions that range from the simple and stable (e.g., sitting) to the complex and unstable (e.g., walking a circuit). Prior accounts of performance in finger identification input may not hold true to this novel setting. To achieve this objective, we calculated metrics for both trial error rate and trial completion time. 
We defined an error trial as one where a finger touched the earbud, causing a magnetic fluctuation and accelerometer impact, but did not trigger the capacitive sensor. Such trials involve initial earbud touches that miss the capacitive sensor and instead impact other nearby areas of the earbud. We defined trial completion time only for successful trials and as the difference between the end of the verbal instructions and the start of a touch to the capacitive sensor. This data, for both metrics and all conditions, is shown in Table~\ref{table_touchbio}. 

Overall, this data showed a grand mean for error rate of 5.6\% (SD: 2.7, representing 484 trials in total) and a mean trial completion time in successful trials of 0.98 seconds (SD: 0.19). All data was normally distributed and upheld sphericity assumptions, so we analyzed both metrics with three-way repeated measures ANOVAs on the variables of posture, finger, and hand. The results have the twin benefits of being both concisely reportable and impactful. Specifically, for trial completion time, there were no significant differences in either interactions or the main effect of hand (with F values between 0.42 and 2.23 and p values in the range of 0.12 to 0.89). There were significant main effects of the finger (F (2, 46) = 4.99, p = 0.011, $\hat{\eta}^2_G$=0.009) and posture (F (2, 46) = 3.33, p = 0.045, $\hat{\eta}^2_G$=0.016), both with small effect sizes. Post-hoc pairwise t-tests using the Benjamini-Hochberg procedure to control the false discovery rate showed a single significant result for posture: trials performed while seated were completed faster than those while standing (in 0.95 seconds vs. 1.01 seconds, p<0.05). For the finger variable, the only significant pairwise result was that trials with the index finger were slower than those with the ring finger (in 1.02 seconds vs. 0.97 seconds, p<0.01). 

For error rate, there were again no significant results for any of the interactions and, additionally, the main effects of posture and finger (with F values between 0.01 and 1.41 and p values in the range of 0.23 to 0.99). The main effect of the hand was significant with a small effect size (F (1, 23) = 4.79, p = 0.039, $\hat{\eta}^2_G$=0.029): trials performed with the left hand achieved a lower error rate than those with the right hand (respectively 4.88\% and 6.32\%). We interpret these results as highly positive: overall, the finger identification input was accurate and rapid, suggesting it would be suitable for real-world use, including in demanding mobility situations such as while walking. While somewhat surprising significant differences did emerge (e.g., faster performance with the ring finger, more accurate performance with the non-dominant hand), the effect sizes and numerical differences were both very small. They are, therefore, unlikely to exert a substantial impact on real-world use of the technique---as such, our objective data supports the general conclusion that bi-manual use of finger identification interaction techniques is both viable and practical on earbuds in diverse typical settings, such as while users are sitting, standing or walking. 

We also analyzed the NASA-TLX scores reported by participants after every combination of posture and hand condition with a two-way repeated measures ANOVA, finding no significant interactions or main effects (F values between 1.76 and 2.11 and p values between 0.13 and 0.19). This suggests there were very limited variations among the conditions in terms of the levels of workload they evoke. The mean scores per item were: mental demand (4.28/20, SD: 4.72), physical demand (4.38/20, SD: 5.09), temporal demand (5.47/20, SD: 5.45), performance achieved (3.42/20, SD: 3.21), effort expended (4.49/20, SD: 4.66), and frustration experienced (2.97/20, SD: 4.07). These scores represent levels of workload typically associated with tasks that are considered undemanding and straightforward ~\cite{Grier15}. However, when asked to embellish this data in relation to the finger variable in post-study rankings of overall preference, index finger taps were preferred (66.67\%), followed by middle taps (33.33\%). Taken together, this subjective data suggests participants experienced the study tasks as simple and that while they did indicate that index finger taps were preferable, the impact of this opinion on empirically measurable aspects of performance was limited. Overall, this data supports our conclusions from the objective data: participants were readily able to tap eyes-free on earbuds with index, middle, and ring fingers in a range of contexts, including while walking.

Finally, we note that participants' pace during the walking conditions in the study was a mean of 1.13 m/s (SD: 0.22), a figure close to the 1.25 m/s mean walking speed of the general population~\cite{alves2020walkability}. This provides further evidence to support our claims that earbud tapping interactions with different fingers are immune to the potentially negative impact of mobility on performance and experience: participants were able to complete our study tasks while walking naturally. This adds further weight to our conclusion that finger identification interactions on earbuds are suitable for real-world use while on the go.
\subsection{Classification Performance}
Encouraged by this strong performance, we turned our attention to developing classifiers that can reliably disambiguate the touching finger. During all classifier development, we excluded the 484 error trials, employed a 6:2:2 data split for training, testing, and validation, and used 5-fold cross-validation procedures. In this analysis, we explored finger classification performance with both deep learning and SVM classifiers. We preprocessed captured data and designed classifiers according to the procedures set out in Section~\ref{section:system}.

\subsubsection{Deep-learning Classifier}
We first explored the performance of our deep-learning classifier in terms of our study variables. To do this, we fixed \textit{nSamples} to 80 (representing 1.33 seconds of data) and trained three posture classifiers, each on all user trials from each pose, and two hand classifiers, each on all user trials with each hand. All models showed broadly similar accuracies for different postures: between 95.3\% (while walking), 96.5\% (while sitting), and 96.6\% (while standing), and for different tapping hands: 96.0\% (right hand) and 97.1\% (left hand). Based on these homogeneous results, we then created three further classifiers. One simply used all magnetometer data, while the other two also included the hand or pose variable as a feature. The results remained similarly homogeneous, ranging from 96.4\% to 96.5\%. Consequently, we opted to conduct all further explorations of classifier performance using the full set of data collected from all conditions in the study. We next considered the use of additional sensor channels. This idea stems from closely related prior work examining finger identification using a ring-mounted magnet on a smartwatch~\cite{geunwoo2020magtouch}. In this prior study, motion data (from accelerometer and gyroscope sensors) was considered critical for achieving high accuracy as it captures changes in a user's orientation that can account for variations in the ambient magnetic field acting on the wearable device. Finger identification systems may need to take such variations into account to achieve good performance. Reflecting the logic from this prior work, we investigated whether accelerometer and gyroscope sensor data can improve our approach to the finger identification problem. We achieved this by processing this additional sensor data in the same way as we treat the magnetometer data and creating a combined classifier. The results were not effective, achieving an accuracy of 95.3\%, a figure slightly worse than that achieved with the magnetometer alone. This suggests that the data about the earbud's motion and rotation contained very little information pertaining to the touching finger and, additionally, that data about variations in the ambient magnetic fields was also of limited value. We corroborated this finding by constructing a model using just accelerometer and gyroscope data, achieving near chance performance of 35.7\%. Based on these outcomes, we opted to use only the magnetometer data for further classifier development.

\begin{table}[t]
\centering
\setlength{\tabcolsep}{4pt}
\Description{This table, titled "Accuracy (\%), shows results from our general model for different data capture periods, expressed in terms of nSamples (the number of samples) and time (seconds). It is divided into two sections: "Before/After Touch" and "Before Touch Only."  In the 'Before/After Touch' section, the table displays accuracy percentages for various combinations of nSamples and time, with an emphasis on high-accuracy cases. Notably, the 40/40 example with a time of 1.33 seconds exhibits high accuracy, achieving an accuracy rate of 93.9\%. The 'Before Touch Only' section showcases accuracy percentages for cases where data is captured only before touch, with nSamples ranging from 30 to 50 and corresponding time values from 0.5 to 0.83 seconds.}
\caption{Accuracy (\%) from our general model for different data capture periods, expressed in terms of \textit{nSamples}, the number of samples and time (seconds).}
        \label{table_gridsearch_time}
        \begin{tabular}{l|ccc|ccc|ccc}
         & \multicolumn{3}{c|}{\textbf{Before/After Touch}} & \multicolumn{3}{c|}{\textbf{Before Touch Only}} & \multicolumn{3}{c}{\textbf{After Touch Only}}\\
        \hline
        nSamples (\#)  & 30/30 & 40/40 & 50/50 & 30/0 & 40/0 & 50/0 & 0/30 & 0/40 & 0/50 \\
        Time (s) & 1 & 1.33 & 1.67 & 0.5 & 0.67 & 0.83 & 0.5 & 0.67 & 0.83 \\
        Accuracy (\%) & 96.0 & 96.4 & 96.4 & 91.8 & 93.9 & 93.8 & 95.1 & 95.6 & 95.3 \\
    \end{tabular}
\end{table}

Following prior work~\cite{kim2022sonarid}, we next explored how the time period in which we collect data impacts the classification accuracy. We did this by varying \textit{nSamples} between 60 (representing a time window of 1 second, centered around the touch mid-point) and 100 (a similarly centered window of 1.67 seconds). We additionally explored half-sized windows truncated at the touch mid-point, including those using data only from before the mid-point and those using data from only after it. The performance of these models is detailed in Table~\ref{table_gridsearch_time}. Peak performance, achieved in the model operating on data before and after a touch and with a window size of 1.33 seconds (\textit{nSamples} = 80), is 96.4\%. In models excluding data before or after the touch point, there is a modest drop in performance, with the best performing model (using a half-sized window of \textit{nSamples} = 40) achieving accuracies of 93.9\% and 95.6\%, respectively, for before- and after-touch data. Although the impact is limited, this result suggests that data from both before and after touch, periods in which the finger will be approaching and retracting from the ear, can help disambiguate which digit initiated contact. Additionally, these results demonstrate effective performance can be achieved under different constraints: low latency (using only data before a touch) and power efficiency (by activating magnetic sensing only after touch). We selected the best-performing model configuration, with data captured from 1.33 seconds around the mid-point of a touch, for all further tests.

\begin{figure*}[t!]
\centering
   \includegraphics[width=\textwidth]{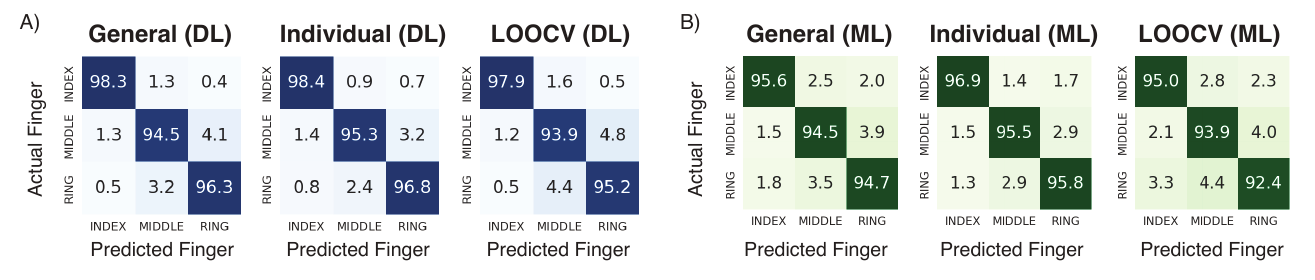}
   \hfil
   \caption{Confusion matrices (\% accuracy for general model and mean \% accuracy for individual and LOOCV models) for deep-learning (A) and machine-learning (B) classifiers.}
   \Description{
   Figure 5 shows a total of six confusion matrices. More specifically, for each deep-learning classifier (A) and machine-learning classifier (B), there are three confusion matrix plots labeled as "General Model," "Individual Model," and "LOOCV Model." In each plot, the x-axis represents the predicted finger, and the y-axis represents the actual finger, with three fingers in consideration: index, middle, and ring. Overall, Deep learning classifier performances modestly outperform machine learning classifiers, and identifying index fingers show better performances (in a range of 97.9\% to 98.3\%) compared to others (in a range of 93.9\% to 96.8\%) in deep learning classifier. Those major confusions appeared between the middle and ring fingers. SVM classifier shows a similar tendency.
   }
	\label{fig:ConfMaxtices}
\end{figure*}

Next, to further validate the system's performance, we generated two additional sets of models: individual models and Leave One Out Cross Validation (LOOCV) models. Individual models, which have been commonly used in prior work~\cite{geunwoo2020magtouch, gil2017tritap}, were generated separately for each participant and trained and tested using only their data. They represent performance levels that could be achieved if genuine users opted to collect data and train their own personalized classifiers. LOOCV models were also generated separately for each participant, but all training data came from the full set of other participants, with the current participant's data only being used for testing. These models represent the level of performance that could be expected with unknown new users. The performance of these models is shown in Table~\ref{table_participantModels}\textcolor{changes}{, and confusion matrices are shown in Figure~\ref{fig:ConfMaxtices}}. Notably, we observed modestly reduced performance for P9, who had the highest self-reported errors, likely due to confusion in tapping fingers leading to greater variations in around-device movements (e.g., switching fingers mid-gesture). While mean individual model accuracy is marginally elevated compared to LOOCV accuracy (at 96.9\% vs 95.6\%), both these results remain broadly in line with the performance of the general model. This is a positive result that we interpret to mean that our system can provide reliable finger classification performance for unseen users as they operate a device in a range of typical poses: on either ear and while sitting, standing, and walking. 

\begin{table*}[t]
\small
\setlength{\tabcolsep}{1.3pt}
\caption{Accuracy (\%) for Individual and LOOCV models on deep-learning classifier}
\Description{
Accuracy (\%) compares individual and Leave-One-Out Cross-Validation (LOOCV) models across 24 participants. Individual model accuracy ranges from a minimum of 90.6\% (Participant 9) to a maximum of 99.4\% (Participants 10 and 16), with a mean accuracy of 96.9\% (SD: 2.1\%). In contrast, the LOOCV model demonstrates accuracy ranging from a minimum of 86.4\% (Participant 9) to a maximum of 99.1\% (Participant 15), with a mean accuracy of 95.6\% (SD: 3.0\%). This table provides insights into the performance variation between individual and cross-validated models across participants.}
\label{table_participantModels}
\begin{tabular}{l|cccccccccccccccccccccccc|cc}
                    & \multicolumn{24}{c}{\textbf{Participant Number}} \\
                    & \textbf{1}    & \textbf{2}   & \textbf{3}    & \textbf{4}     & \textbf{5}    & \textbf{6}      & \textbf{7}   & \textbf{8}    
                    & \textbf{9}	 & \textbf{10}	 & \textbf{11}	  & \textbf{12}	& \textbf{13}	 & \textbf{14}	    & \textbf{15}  & \textbf{16} & \textbf{17}   & \textbf{18}    & \textbf{19}	 & \textbf{20} & \textbf{21}    & \textbf{22}   & \textbf{23}    & \textbf{24} & \textbf{$\mu$} & \textbf{$\sigma$} \\
\hline
Indiv.     & 94.9 & 98.7 & 92.8 & 98.9 & 96.4 & 96.6 & 94.2 & 96.1 & 90.6 & 99.4 & 98.8 & 97.6 & 98.5 & 98.8 & 98.0 & 99.4 & 96.2 & 94.6 & 96.1 & 97.7 & 97.1 & 97.1 & 97.9 & 98.2 & 96.9 & 2.1 \\
LOOCV    & 93.7 & 97.5 & 89.3 & 98.9 & 97.3 & 91.1 & 94.4 & 97.9 & 86.4 & 96.5 & 98.0 & 97.9 & 98.5 & 97.9 & 99.1 & 95.3 & 96.7 & 97.0 & 95.1 & 95.0 & 95.3 & 97.4 & 96.4 & 93.5 & 95.6 & 3.0\\
\end{tabular}
\end{table*}

\subsubsection{SVM classifier}
We assessed the performance of SVM classifiers using the optimal configuration identified in our explorations with the deep learning classifier. Specifically, we combined all study data, set \textit{nSamples} to 80, and trained general, individual, and LOOCV models. The results revealed minor reductions in accuracy of between 0.8\% (mean individual models) to 1.9\% (mean LOOCV models). These tended to be due to lower accuracy for classifying index finger touches---see Figure~\ref{fig:ConfMaxtices}-B for full details. This is likely due to the statistical features used in the SVM classifiers failing to capture the full nuances of performance present in the raw data used with the deep learning classifiers. Regardless, performance remains high, and the simple features and low computational costs of SVM classification highlight the practicality and suitability of our approach for low-resource wearable devices. 
\section{Study 2: Consecutive Double-Tap Classification}
Building on the positive results of our first study, we conducted a second study to explore the design of more expressive techniques for finger identification on earbuds. Specifically, we focused on the idea of \textit{multi-finger double-tap}, or the use of pairs of rapidly sequential touches made by different fingers. We note that while sequential touches have been frequently proposed to augment interaction on small form factor devices such as smartwatches~\cite{Spatial2016Benjamin, zoomboard13stephen} and earbuds~\cite{xu2020earbuddy}, our multi-finger variant is novel. This study was approved by the local IRB.

\subsection{Participants}
We recruited 16 participants (9 males) with a mean age of 23.13 (SD 2.36) through an online student community. Among them, 14 were right-handed, 1 was left-handed, and 1 was ambidextrous. The participants had an average hand length of 17.79 cm (SD 0.94) and a palm width of 7.79 cm (SD 0.39). The average length of their index, middle, and ring fingers was 7.13 cm (SD 0.34), 7.74 cm (SD 0.36), and 7.08 cm (SD 0.45), respectively. The study took approximately one hour, and participants received the equivalent of 15 USD in local currency as compensation for their participation.

\subsection{Design and Procedures}
This study featured two simple independent variables: first and second touching fingers. Each had three levels (index, middle, and ring) leading to nine different conditions. For each condition, we captured data from 40 repetitions for each participant (360 trials in total). To minimize fatigue, we divided these trials into eight equal blocks of 45 trials, each separated by a rest period of one minute. Half the blocks were conducted using the left buds and half with the right buds. Although we did not consider handedness a formal variable in this study (due to its low impact in the first study), we balanced participants such that half of them completed all left trials before right trials and the others vice versa. Finally, each block of 45 trials was structured to include 5 repetitions of each of the nine conditions, with each trial delivered in a random order. In this way, we captured 5760 trials in this study (9 conditions by 5 repetitions by 8 blocks by 16 participants). 

This study followed broadly similar procedures to the first study: obtaining consent, donning equipment, delivering instructions, practicing, and then finally executing the study. It was again conducted in a quiet classroom environment, but as our focus was input expressivity, participants were seated in front of a desk throughout. In addition, they were encouraged to rest their elbow on the desk to reduce fatigue. Instructions in the study were again delivered verbally but came as pairs of ordered finger combinations (e.g., "index and middle"). Practically, in order to detect a double-tap (and separate it from a sequence of single taps), we also need to establish a maximum temporal separation between pairs of touches. On desktop computers, this is commonly 500ms for mouse double click\footnote{https://learn.microsoft.com/en-us/windows/win32/controls/ttm-setdelaytime}. However, prior work on sequential touches involving different fingers~\cite{FingerText21Lee} reports somewhat elevated inter-touch times of 583ms. Accordingly, in this study, we set an intentionally generous maximum inter-touch time of one second. We provided verbal feedback ("missed") and considered the trial an error if the gap between participants' taps exceeded this duration. 
\subsection{User Performance Results}
Participants self-reported a total of 52 errors (0.9\%) in this study, and we additionally noted 44 trials suffering technical failures (0.76\%). We excluded these trials to retain data from 5664 trials for analysis. We then calculated and analyzed time and error metrics. Errors were defined as in the first study, and we additionally included trials in which the inter-touch duration exceeded one second. Time metrics, calculated only for successful trials, included touch time (the duration between delivery of the verbal instruction and first contact with the sensor) and inter-touch time (the time between the start points of the two touches in each trial). Overall, participants committed 160 errors (mean error rate: 2.8\%, SD: 1.7), a rate notably reduced from the 5.6\% recorded in the first study. This may be due to variations in the experimental procedure: in the first study, participants raised and lowered their hands, while in this study, their hands remained in a fixed posture near the ear. We suggest increased targeting errors may occur when the hand is initially and rapidly raised to the ear. Regarding the time data, participants achieved a mean time of 0.94 seconds (SD: 0.18), in line with the first study, and a mean inter-touch time of 0.40 seconds (SD: 0.08), below the 0.5s threshold used for typical double clicks. These data indicate participants were able to complete multi-finger double-taps accurately and rapidly.  

\begin{figure*}[t!]
\centering
   \includegraphics[width=15cm]{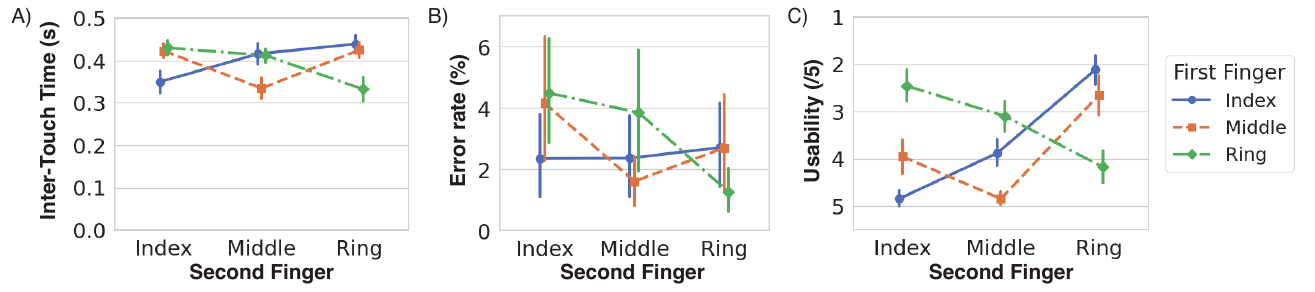}
   \hfil
   \caption{Interaction plots of inter-touch time (A), error rates (B), and self-reported usability (measured with 5-point Likert Scale) (C) for different sequential taps of first and second fingers.}
   \Description{Figure 6 illustrates interaction plots of inter-touch time (A), error rates (B), and self-reported usability (measured with a 5-point Likert Scale) (C) for different sequential taps of first and second fingers. Figure 5-(A) shows that reduced inter-touch times (going from 0.41s-0.44s to 0.33s-0.35s) occur when a pair of taps involve the same finger (e.g., index-index or ring-ring). Figure 5-(B) indicates that ring finger second touches showed reduced errors compared to index finger second touches (respectively 2.21\% and 3.66\%), an effect likely due to the extremely low error rates recorded for repeated touches of the ring finger (1.3\%). Lastly, Figure 5-(C) suggests participants strongly favored repeated taps with the same finger. Participants rated taps involving the ring finger poorly compared to taps involving the middle and index fingers.}
   \label{fig:doubletap}
\end{figure*}

All data was normally distributed and upheld sphericity assumptions. We thus conducted two-way repeated measures ANOVAs on the variables of first and second touching fingers. Touch-time showed no significant differences for either variable (with F values ranging from 0.05 to 1.99 and p-values between 0.14 to 0.99) while inter-touch time showed a significant interaction effect (F(4, 8) = 40.1, p < 0.001, $\hat{\eta}^2_G$=0.33) and main effect of second finger (F(2,14) = 4.8, p=0.015, $\hat{\eta}^2_G$=0.01). We then plotted the interaction effects (see Figure~\ref{fig:doubletap}) and conducted post-hoc pairwise t-tests using the Benjamini-Hochberg adjustments on the main effect. The interaction plots show that reduced inter-touch times (going from 0.41s-0.44s to 0.33s-0.35s) occur when a pair of taps involve the same finger (e.g., index-index or ring-ring). Further, the main effect of the second finger was due to a small reduction (from 0.4s to 0.38s) when the middle finger was employed compared to the index and ring (both p<0.017). These effects likely reflect the reduced hand travel distances required for double-tap with the same finger and middle finger second touches. The only significant effect for error rate was for second finger touches (F(2,14) = 3.6, p=0.041, $\hat{\eta}^2_G$=0.032). Post-hoc analysis indicated that ring finger second touches showed reduced errors compared to index finger second touches (respectively 2.21\% and 3.66\%, p=0.028), an effect likely due to the extremely low error rates recorded for repeated touches of the ring finger (1.3\%). 

At the end of the study, participants provided ratings (on 5-item Likert scales) for each pair of finger taps using four metrics: preference, comfort, speed, and confusability. Responses were highly correlated (mean Pearson's r(14)=0.97), so we analyzed their mean, again using a two-way ANOVA on the variables of first and second touching finger. The interaction effect (F(4,8) = 125.6, p<0.001, $\hat{\eta}^2_G$=0.62) and main effects of first (F(2,14) = 23.3, p<0.001, $\hat{\eta}^2_G$=0.30) and second (F(2,14) = 12.2, p<0.001, $\hat{\eta}^2_G$=0.12) finger were all significant. The interaction effects are shown in Figure~\ref{fig:doubletap}-C; examination of the plot suggests participants strongly favored repeated taps with the same finger. Post-hoc pairwise t-tests on the main effects indicate participants rated taps involving the ring finger poorly (all p<0.004) compared to taps involving the middle and index fingers. Finally, participants completed NASA-TLX at the end of the study to provide an overall assessment of their experience. Mean scores for each NASA-TLX item were: mental demand (5.13/20, SD: 4.92), physical demand (5.5/20, SD: 4.73), temporal demand (4.88/20, SD: 4.23), performance achieved (3.63/20, SD: 1.68), effort expended (5.53/20, SD: 4.84), and frustration experienced (1.88/20, SD: 2.39). While these show a modest increase over the scores in the single-tap study (respectively, 4.43/20 vs 4.16/20), they still represent workload levels associated with undemanding and straightforward tasks~\cite{Grier15}. We conclude that, despite relatively poor ratings for ring finger taps, our multi-finger double-tap interaction technique did not place undue burdens on participants.  
\begin{figure*}[t!]
\centering
   \includegraphics[width=16cm]{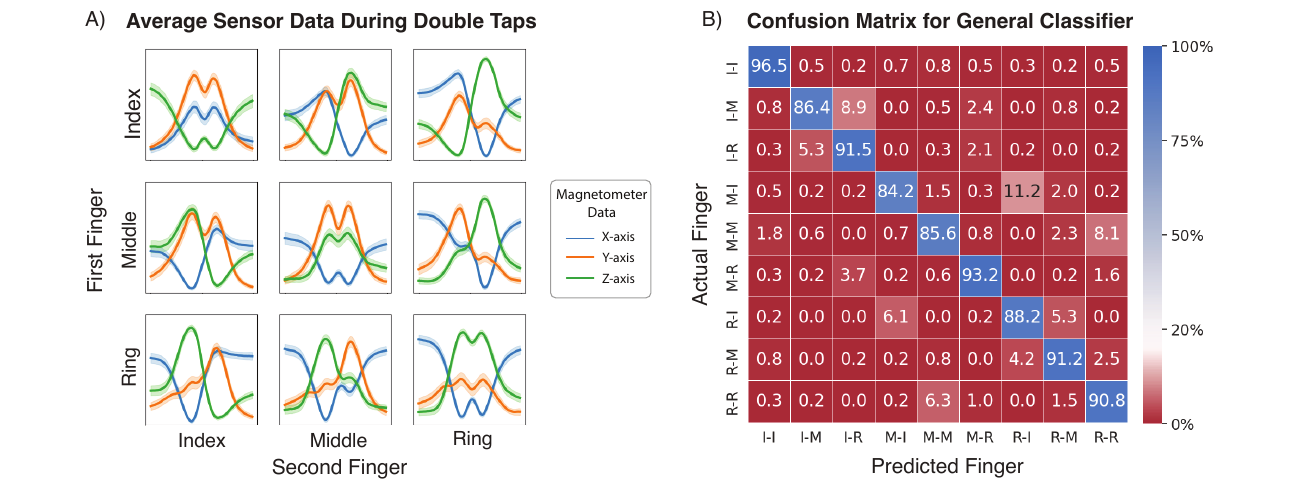}
   \hfil
    \caption{
    (A) Visual depiction of magnetometer sensor data during double-taps. (B) Confusion matrix for the general model of 1D-CNN classifier. Touching fingers are denoted by their initials (e.g., I-M signifies an \textit{I}ndex followed by a \textit{M}iddle finger touch).
    }
    \Description{
    In Figure 7, the left image (A) depicts the overall visual examination of magnetometer sensor data during double-taps showing clear double peaks (crest or trough). Similar to study 1, index finger taps show a crest in the X-axis data and a trough in the Z-axis data, whereas the opposite pattern is observed for the middle and ring fingers. Additionally, while ring taps exhibit modestly greater variability in both X and Z axes compared to Y-axis data, middle taps show relatively low variability in X and Z axes compared to larger variability in Y data.
    }
	\label{fig:doubletapConf}
\end{figure*}

\subsection{Classification Performance}
Building on these positive results, we explored classification performance for the nine different inputs enabled by our technique. Due to this increased number of classes, the number of samples available per class was reduced. Consequently, we opted to increase the proportion of data used for training. We employed an 8:1:1 data split for training, testing, and validation (for deep learning) and 10-fold-cross validation procedures (for machine learning). All data were again preprocessed following the procedures outlined in Section~\ref{section:system}. To accommodate the longer tasks involved in repeated taps, we set \textit{nSamples} to 100 (up from 80), a 0.33s increase that roughly corresponds to the mean measured inter-touch time. Figure~\ref{fig:doubletapConf}-A displays the mean of all captured magnetometer data in our second study for all nine combinations of first and second finger touches. We note data again shows highly consistent and visually distinct patterns, albeit with unsurprisingly increased complexity. 

We first explored performance in identifying consecutive tapping fingers using our 1D-CNN classifier. The general model achieved an accuracy of 91.3\%; Figure~\ref{fig:doubletapConf}-B shows the resulting confusion matrix. While we note performance remains good, especially given the increased number of classes that must be distinguished, we also acknowledge the presence of some more frequently confused classes. These, in general, involve confusion between the middle and ring fingers. For example, the most commonly occurring misclassification (at 11.2\%) involves returning 'R(ing)-I(index)' for 'M(iddle)-I(ndex)': the middle finger is mistaken for the ring, while the index finger is correctly recognized. This trend continues for other elevated errors and also matches data observed in the first study. We next fleshed out our performance characterization by generating individual and LOOCV models, attaining mean accuracies of, respectively, 94.7\% (SD: 1.4\%) and 82.4\% (SD: 7.4\%). While individual model performance is on par with the first study, the notable drop in performance in LOOCV models may be due to the relatively small size of the training data available for each class (decreasing from 120 samples to 40 samples per participant per class). 

We then applied a modified version of the SVM classifier. We retained \textit{nSamples} at 100, but divided each data window into two equal halves, temporally centered at the mid-point of the pair of touches, and calculated the statistical feature set (six features for each of the three sensor axes) used in the first study separately for each half. This resulted in 36 features, with 18 derived from data captured during the first touch of each pair and 18 based on data from the second touch. The results show reduced classification accuracy compared to 1D-CNN classifier across the board: in the general model (82.6\%), the individual models (a mean of 89.5\%, SD: 4.6) and, in particular, in the LOOCV models (a mean of 68.9\%, SD: 10.1). This reduced performance suggests that simple statistical features may be inadequate to capture the more complex patterns in the data that occur during multi-finger double-tap input.
\section{BudsID Applications}
Given these strong usability and classification accuracy results, we explored and evaluated designs for earbud applications that use the full range and potential of the finger identification interaction techniques we propose. Figure~\ref{fig:application} illustrates possible examples of interface designs using finger identification to enhance earbud use.

\subsection{Interaction Design}

\subsubsection{Contextual Input}
One simple, effective approach would be \textit{contextual}--it would associate different functionality with different fingers depending on the current state of the device. For example, when receiving a call, single taps by different fingers could be associated with accepting, switching, or declining it. Conversely, when listening to music, single taps could navigate to prior and subsequent songs in a playlist and toggle between playing and pausing the music.

\subsubsection{Hierarchical Input}
Finger identification input responses could be seamlessly integrated (with a simple temporal threshold, as with single/double clicks on a mouse) with more complex \textit{hierarchical} input based on rapidly issued pairs of taps. For example, an initial tap with the index finger could access a set of three further commands relating to volume control (e.g., raising, lowering, toggling mute) that are selected by a follow-up tap. Similarly, starting a pair of taps with the middle finger could open up a set of commands related to Bluetooth device connections, and initial ring taps could access yet further functionality (e.g., turning on the lights or controlling IoT devices in the home). There is precedent for such functionality as prior researchers have proposed various systems~\cite{Beats2015Ian, Spatial2016Benjamin} based on pairs of rapid touches to wearable devices such as smartwatches. Additionally, prior work has explored the viability of such complex inputs on eyes-free devices, ultimately suggesting that carefully designed audio feedback~\cite{earpod07shengdong} can effectively support users in navigating menus. 

\begin{figure}[t!]
\centering
   \includegraphics[width=\textwidth]{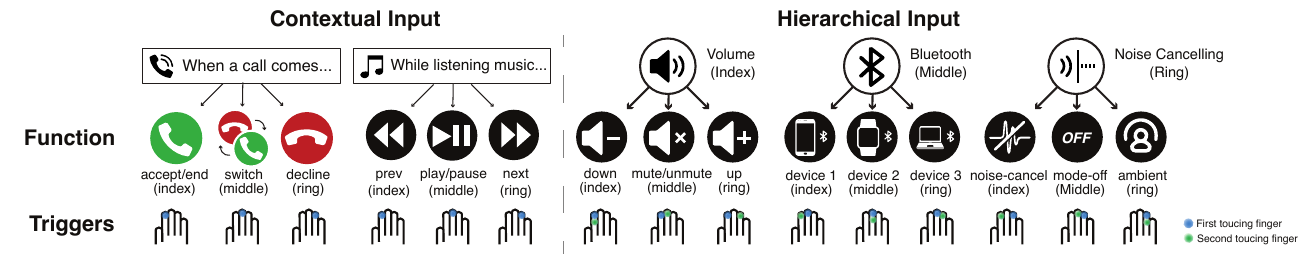}
   \hfil
\caption{Example application functions \textcolor{changes}{annotated with an illustration of the fingers used during operation of BudsID for both contextual (left) and hierarchical input (right)}. 
}
\Description{
Figure 8 illustrates key example applications for finger identification on earbuds showing contextual input (left), and hierarchical input (right). In a contextual input scenario, firstly, when receiving a call, single touches by different fingers could be associated with accepting, switching, or declining it. next, when listening to music, single touches could navigate to prior and subsequent songs in a playlist, and toggle between playing and pausing the music. In a hierarchical input scenario, the initial tap with the middle finger could access a set of three further commands relating to volume control such as raising, lowering, and toggling mute that are selected by the follow-up tap. Similarly, starting a pair of taps with the middle finger could open up a set of commands related to Bluetooth device connections and initial ring taps could access yet further functionality.
}
\label{fig:application}
\end{figure}

\subsection{User Feedback Study}
To gather feedback on our BudsID designs, we conducted a user study where participants compared our system and a set of baseline input techniques using the finger tap gestures common on commercial earbuds~\footnote{https://www.samsung.com/us/support/answer/ANS10001318/} such as single tap to play/pause music, double tap to skip tracks, and triple tap for playing prior tracks. We also extended current designs by applying this existing input repertoire to novel tasks such as Bluetooth pairing by, for example, using a long tap to enter pairing mode and single, double, and triple taps to manage device connections. Consequently, this study evaluated five different scenarios, as shown in Figure~\ref{fig:application}, including two contextual and three hierarchical input scenarios. This study aimed to observe and learn from novice users' comments and reactions regarding our system. It was approved by the local IRB. 

\subsubsection{Participants}
We recruited 10 participants (5 males, mean age of 24 (SD 2.4)) from the local university. They were highly familiar with Bluetooth earbuds (4.3/5, SD 0.67) but less familiar with interaction gestures on earbuds (3.3/5, SD 1.34). None had completed the previous single and double tap studies. The study took approximately one hour per participant, and each was compensated with 15 USD in local currency. 

\subsubsection{Procedure}
Participants first signed an informed consent form, and we explained the BudsID system. After donning the equipment, they were given 3 minutes to familiarize themselves with the finger identification feature while seated, and then the study commenced. We first demonstrated the five different use cases (e.g., call, music, volume, Bluetooth, and noise-canceling mode control) with one of the interaction systems (either BudsID or the baseline). For each use case, participants then enacted the scenario for one minute with visual instructions describing the input and spent another minute performing the task with these instructions. At the end of each interaction scenario, participants reported the perceived usefulness, ease, and memorability of the interaction on three 5-point Likert scales, scored from 1 (very unsatisfied) to 5 (very satisfied). After completing all five scenarios, participants reported subjective usability using the System Usability Scale (SUS)~\cite{brooke1996sus}. The participant then repeated this procedure for the other interaction system. The order in which these two systems were experienced was fully balanced across participants. 

\begin{figure}[t!]
\centering
   \includegraphics[width=0.99\textwidth]{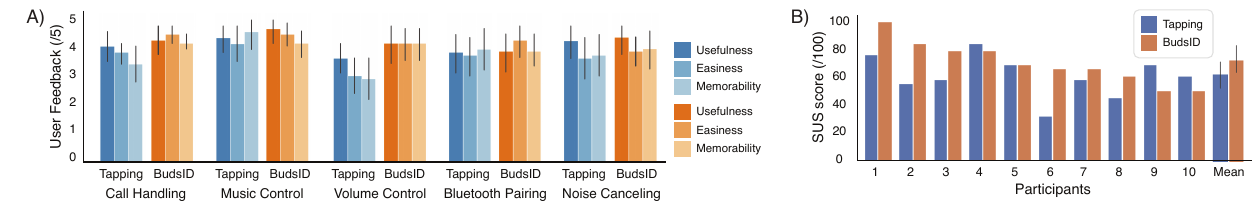}
   \hfil
\caption{Mean subjective rankings for each interaction scenario A), and system usability scores for each participant B).}
\Description{
Figure 9 illustrates a bar chart describing subjective rankings in terms of usefulness, easiness, and memorability for each interaction scenario in the left figure (A) and measured system usability scales for each participant in the right figure (B).
}
\label{fig:study3_result}
\end{figure}

\subsubsection{Results}

Overall, participants provided positive feedback on BudsID. They rated the traditional tapping interaction 3.98 for usefulness, 3.64 for ease of use, and 3.68 for memorability out of 5. In comparison, BudsID received higher ratings: 4.3 for usefulness, 4.28 for ease of use, and 4.06 for memorability. Additionally, BudsID achieved a System Usability Scale (SUS) score of 69.25 (SD 14.8), surpassing the benchmark mean score (commonly reported to be 68~\cite{lewis2018item}), whereas the commercial tapping interaction scored markedly lower at 60 (SD 14.3). A summary of subjective feedback for each interaction scenario and the measured usability scales for each participant are illustrated in Figure~\ref{fig:study3_result}. Key observations from the study are outlined below.

First, participants liked the \textit{simplicity} of BudsID input compared to traditional controls, which can be cumbersome with extra long-taps or triple-taps. For example, our finger identification interaction was considered "much more comfortable" than traditional long-tap (P3, P2, P4, P8) or triple-tap (P1, P2, P7, P8), and dubbed as "simple and handy" (P1, P2, P3, P7). Second, many participants found our finger identification controls to be more "intuitive" (P7). They were seen as more "systematic" (P2, P3) and "memorable" (P1, P2, P3, P5, P6, P7, P10) compared to commercial interactions. P7 summarized this well: "The BudsID interaction is more intuitive and easier to remember compared to traditional methods. While commercial interactions are more complex and involve various patterns such as multi-tap and long-tap, BudsID offers a simpler and more consistent set of rules, making it easier for users to learn and use". Lastly, participants highlighted the effectiveness of our multi-finger interactions, particularly in specific use cases such as driving. For example, P3 mentioned that "This simplified interaction would be much better for those who are driving." and P2 expressed that "With commercial earbuds, I often resort to using my phone for call and music control to avoid multiple taps, but with BudsID interaction, I wouldn’t need to do that as I could easily manage it directly through the earbuds." On the other hand, two participants stated they preferred to use single finger commercial earbud interaction (P9, P10), predominantly due to their high levels of familiarity with those systems. For instance, P9 mentioned difficulty in adapting to new interactions, "I have frequently used tapping interaction of commercial earbuds, so BudsID interaction was more difficult to me as it is something new." Also, some participants expressed inconvenience in wearing the magnet ring. For instance, P3 suggested a smaller or attachable (e.g., to nail) form factor, "Adding an extra accessory like a ring might be inconvenient for daily use. It would be useful if it were smaller or attachable." This suggests a design challenge for our system: making our system more practical for real-world use by improving its size and form factor.

\section{Discussion}
The paper presents BudsID, an earbud system that senses magnetic fluctuations evoked by a magnetic ring worn on the middle finger to determine which fingers (among index, middle, or ring) are being used to perform single or double-taps to the device. While this type of finger identification input has been widely studied in devices such as smartphones~\cite{le2019investigating} and smartwatches~\cite{geunwoo2020magtouch, gil2017tritap, kim2022sonarid}, its adaptation to the out-of-sight form factor of earbuds is novel. In addition, compared to alternative interaction techniques, such as verbal interaction, which are impractical in public settings (e.g., multi-party conversation)~\cite{easwara2014voice, efthymiou2016evaluating} or touch screens that require visual attention and can be distracting (e.g., while walking or driving)~\cite{FELD2019219, touchLook08Ba}, our technique uses simple finger touches, offering direct, straightforward interaction even when eyes-free and mobile. Our final study shows the practical implications of finger identification input for enhancing earbud use in real-world scenarios and applications.

\subsection{Behavioral Performance Data}
Our performance results show considerable promise. Specifically, our studies indicate that participants can rapidly make initial (in 0.98 seconds) and follow-up (in 0.4 seconds) touches while also committing few errors (between 2.8\% and 5.6\%) and reporting low levels of workload. In addition, our first study demonstrates that this high level of performance is reliable in a variety of poses (e.g., sitting and standing) and while walking. Our second study complements this by showing how double-taps can extend the expressivity of finger identification techniques to support a wider range of inputs without compromising performance. These strongly positive results suggest finger identification interaction techniques are a good fit for earbud platforms---they are based on an input task users can perform adeptly in various circumstances and support diverse inputs. We note that these findings are independent of any particular sensing technique for detecting fingers: they relate to, and validate, user performance in the basic task. 

\subsection{Finger Identification Performance}
In addition to these generally applicable results characterizing human performance during multi-finger earbud input, our work also sheds light on the nuances of deploying the magnetic sensing technique we used to enable our work. We apply a previously unstudied approach of using a 1D-CNN to perform classification and show that it can attain high accuracies of 96.4\% and 91.3\% classifying, respectively, single and double-taps in general models. \textcolor{changes}{Both studies also show consistent trends: the highest classification performance is achieved with the index finger, while the middle and ring fingers are more frequently confused. These findings have several implications. Firstly, they suggest that a two-class task differentiating between index taps and taps by middle and ring fingers would achieve improved accuracy (of 97.9\% or higher): designing a system with two basic input events, akin to left and right mouse buttons, would thus achieve very levels high of accuracy. Secondly, we should seek to account for the lower performance between the middle and ring fingers. Possible explanations include the anthropometric, such as the colloquial observation that the interdigital web space (or gap) between the index and middle fingers tends to be larger than the one between the middle and ring fingers~\cite{interdigitalWeb19}. Alternatively, from a behavioral perspective, we observed index finger touches were conducted with the forearm almost vertical, while middle and ring touches more frequently involved forearm rotations modestly toward the horizontal. Such large-scale arm rotations were relatively diverse both within and between participants and may have contributed to the increased difficulty of separating these two classes. Providing explicit instructions~\cite{gil2017tritap} to users on how to touch to achieve the best performance (e.g., to splay fingers, or maintain the forearm in a vertical pose) may ameliorate these variations and boost accuracy. Alternatively, different ring placements might also boost performance---a magnet on the index or ring finger instead of the central middle finger might accentuate differences, as each finger would be a different absolute distance from the magnet. Future work should explore the effectiveness of these ideas. }

We also demonstrate that single-tap performance is robust to variations in pose, orientation, and the disturbances inherent in walking. Furthermore, elevated individual model accuracy with double-taps (of 94.7\%) suggests that modestly larger training set sizes will further boost performance. Finally, we show promising performance in LOOCV models (of 95.6\% (single-tap) and 82.4\%(double-tap)), suggesting it should be practical to deploy the technique to new users. \textcolor{changes}{We contrast these results with performance in closely related smartwatch studies~\cite{geunwoo2020magtouch, gil2017tritap, kim2022sonarid}. These comparisons are broadly favorable: our system is relatively effective for classifying the fingers involved in single-tap with 96.9\% in per-user models (vs. 93.7\% (sonar sensing~\cite{kim2022sonarid}),  95.03\% (Magnetic sensing with SVM classifier~\cite{geunwoo2020magtouch}), and 79.4\% (capacitive touch sensing~\cite{gil2017tritap}). Further, we demonstrate this strong performance holds in more challenging scenarios, such as while walking, apply it to a novel, rapid consecutive double-tap scheme, and illustrate how it could be deployed in application and interaction design.} Overall, we argue the level of performance we report with our 1D-CNN model, achieved in a broad range of scenarios, demonstrates that our approach is sufficiently accurate to support the implementation of diverse input techniques and applications.

\subsection{Deployment}
We additionally explore the technical feasibility of deploying our finger identification classifiers to wearable devices by developing versions of our deep learning classifiers that run on the low-resource microcontroller in our prototype. To determine the viability of our technical approach for real-world deployment, we optimized our general model for microcontroller deployment in our prototype using TinyML~\cite{warden2019tinyml}. The original models have 168.5 to 210K parameters and require 658 to 819 KB of storage, respectively, for single-tap (\textit{nSamples}: 80) and double-tap (\textit{nSamples}: 100) scenarios. By applying dynamic range quantization procedures that convert model weights from high-precision floating point representations to byte representations, we lowered the model size by 3.89 times to, respectively, 166 KB and 207 KB, footprints more suitable for an embedded system. We note that despite this reduced size, general models maintain good performance: 96.0\% for single-taps and 89.1\% for double-taps, indicating quantization has limited impact on model performance. We also note that classification times on the Arduino are appropriate: feature extraction and preprocessing were near instantaneous at 0.16ms and very stable (SD: 0.00), and prediction times were 39.38ms (SD: 14.28), fast enough to support returning results in periods that appear instantaneous to a user. This implementation and evaluation demonstrate that training a lightweight deep learning classifier is feasible even on a low-resource device such as a microcontroller embedded in a pair of earbuds. We include code (and model) for this system in the supplementary files alongside this submission. 

\subsection{Practicability of Additional Magnet Ring}
\textcolor{changes}{Despite the promising potential of our system for increasing input expressivity for earbuds, including a magnet-integrated ring may pose inconveniences for users. Therefore, it is worth discussing the further benefits of integrating such a ring. Practically, a passive magnetic element requires no power or electronics. Our observations (see Section~\ref{fig:5fingers}) also demonstrate the feasibility of reducing the magnet size to a minimal, unobtrusive form factor (of 5mm by 4mm, approximately the size of a typical 0.5-carat center stone\footnote{https://www.diamdb.com/carat-weight-vs-face-up-size/}) that could be relatively easily integrated into a passive or smart ring form factor. Additionally, integrated magnets could be designed to support complementary functions, such as achieving secure connections to smart ring chargers, or could be made to be detachable or swappable, allowing users to customize a ring to their taste. A magnet-integrated ring could also extend its utility beyond earbuds, enhancing interaction opportunities~\cite{Ashbrook11Nenya} with the diverse other devices explored in prior works, such as smartwatches~\cite{geunwoo2020magtouch}, smartphones~\cite{MagRingMobileMHCI19, TowardsUsing10MHCI}, and head-mounted displays (HMDs)~\cite{ThumbRingMHCI16}. In this way, integrating a magnet into the emerging all-day wearable platform of a smart ring could enhance the expressivity of a wide range of other wearables, including other devices a user may wear continuously (e.g., watches or glasses) and those they may wear more sporadically (e.g., earbuds and HMDs), thus moving towards a future with enhanced near-scale cross-device interaction~\cite{CrossDeviceTaxonomyCHI19}.}

\subsection{Toward Expressive Input on Earbuds}
We explored a single input action: tapping. While this is arguably the most basic input primitive, it also represents only a subset of potential interactions a user might employ. First, our approach can be further refined by combining it with complementary techniques for finger identification, such as sonar~\cite{kim2022sonarid} or capacitive sensing~\cite{gil2017tritap}. This may further increase the accuracy and robustness of performance. Beyond adapting these prior approaches, it may be possible to explore alternative forms of instrumentation for the touching finger. For example, rather than directly instrumenting the finger, it may be possible to augment the wrist by adding a passive magnetic buckle to a smartwatch (in place of the magnetic ring) or rely on signals transmitted from~\cite{hessar2016enabling} or sensed on a watch~\cite{listen2020chen} to identify the touching finger. 
\textcolor{changes}{Second, finger identification can be combined with more diverse gestures, such as dwelling, dragging, and swiping~\cite{Goguey16} to enhance the expressiveness of the approach. Around-ear interaction studies have proposed a wide range of input gestures in different spaces, including on-face~\cite{xu2020earbuddy, rateau2022leveraging}, on the ear lobe~\cite{kikuchi2017eartouch, xu2020earbuddy}, and back of the ear~\cite{lissermann2014earput, xu2020earbuddy}. Our system could expand the capabilities of these interaction spaces by distinguishing the interacting finger. For example, EarBuddy~\cite{xu2020earbuddy} classified tapping and sliding gestures near the face and ears using a microphone. Integrating BudsID into this system could enhance its performance by identifying which finger performs the interaction, thereby enabling more nuanced input capabilities. Future work should investigate how these potential enhancements translate to real-world usability and acceptance. 
} 
\section{Conclusion}
In conclusion, this paper is the first to demonstrate the practicality and feasibility of finger identification on earbuds. Our participants' ability to tap reliably on earbuds was well matched by BudsID's ability to accurately identify the touching finger, even in challenging situations such as while walking and during rapid sequential double-taps. Based on this combination of strong input and accurate classification performance, we argue that the finger identification input paradigm is well matched to the earbuds form factor. To emphasize this point, we close by presenting and evaluating the design of practical applications using finger identification that highlight the rich range of input possibilities it enables on earbuds.


\bibliographystyle{ACM-Reference-Format}
\bibliography{sample-base}


\begin{thebibliography}{72}


\ifx \showCODEN    \undefined \def \showCODEN     #1{\unskip}     \fi
\ifx \showDOI      \undefined \def \showDOI       #1{#1}\fi
\ifx \showISBNx    \undefined \def \showISBNx     #1{\unskip}     \fi
\ifx \showISBNxiii \undefined \def \showISBNxiii  #1{\unskip}     \fi
\ifx \showISSN     \undefined \def \showISSN      #1{\unskip}     \fi
\ifx \showLCCN     \undefined \def \showLCCN      #1{\unskip}     \fi
\ifx \shownote     \undefined \def \shownote      #1{#1}          \fi
\ifx \showarticletitle \undefined \def \showarticletitle #1{#1}   \fi
\ifx \showURL      \undefined \def \showURL       {\relax}        \fi
\providecommand\bibfield[2]{#2}
\providecommand\bibinfo[2]{#2}
\providecommand\natexlab[1]{#1}
\providecommand\showeprint[2][]{arXiv:#2}

\bibitem[Ahuja et~al\mbox{.}(2021)]%
        {PilotEar21Ferlini}
\bibfield{author}{\bibinfo{person}{Ashwin Ahuja}, \bibinfo{person}{Andrea Ferlini}, {and} \bibinfo{person}{Cecilia Mascolo}.} \bibinfo{year}{2021}\natexlab{}.
\newblock \showarticletitle{PilotEar: Enabling In-ear Inertial Navigation}. In \bibinfo{booktitle}{\emph{Adjunct Proceedings of the 2021 ACM International Joint Conference on Pervasive and Ubiquitous Computing and Proceedings of the 2021 ACM International Symposium on Wearable Computers}} (Virtual, USA) \emph{(\bibinfo{series}{UbiComp/ISWC '21 Adjunct})}. \bibinfo{publisher}{Association for Computing Machinery}, \bibinfo{address}{New York, NY, USA}, \bibinfo{pages}{139–145}.
\newblock
\showISBNx{9781450384612}
\urldef\tempurl%
\url{https://doi.org/10.1145/3460418.3479326}
\showDOI{\tempurl}


\bibitem[Alves et~al\mbox{.}(2020)]%
        {alves2020walkability}
\bibfield{author}{\bibinfo{person}{Fernando Alves}, \bibinfo{person}{Sara Cruz}, \bibinfo{person}{Anabela Ribeiro}, \bibinfo{person}{Ana Bastos~Silva}, \bibinfo{person}{João Martins}, {and} \bibinfo{person}{Inês Cunha}.} \bibinfo{year}{2020}\natexlab{}.
\newblock \showarticletitle{Walkability Index for Elderly Health: A Proposal}.
\newblock \bibinfo{journal}{\emph{Sustainability}} \bibinfo{volume}{12}, \bibinfo{number}{18} (\bibinfo{year}{2020}), \bibinfo{numpages}{27}~pages.
\newblock
\showISSN{2071-1050}
\urldef\tempurl%
\url{https://doi.org/10.3390/su12187360}
\showDOI{\tempurl}


\bibitem[Ashbrook et~al\mbox{.}(2011)]%
        {Ashbrook11Nenya}
\bibfield{author}{\bibinfo{person}{Daniel Ashbrook}, \bibinfo{person}{Patrick Baudisch}, {and} \bibinfo{person}{Sean White}.} \bibinfo{year}{2011}\natexlab{}.
\newblock \showarticletitle{Nenya: Subtle and Eyes-Free Mobile Input with a Magnetically-Tracked Finger Ring}. In \bibinfo{booktitle}{\emph{Proceedings of the SIGCHI Conference on Human Factors in Computing Systems}} (Vancouver, BC, Canada) \emph{(\bibinfo{series}{CHI '11})}. \bibinfo{publisher}{Association for Computing Machinery}, \bibinfo{address}{New York, NY, USA}, \bibinfo{pages}{2043–2046}.
\newblock
\showISBNx{9781450302289}
\urldef\tempurl%
\url{https://doi.org/10.1145/1978942.1979238}
\showDOI{\tempurl}


\bibitem[Ates et~al\mbox{.}(2015)]%
        {cagri2015expanding}
\bibfield{author}{\bibinfo{person}{Halim~Cagri Ates}, \bibinfo{person}{Ilias Apostolopoulos}, {and} \bibinfo{person}{Eelke Folmer}.} \bibinfo{year}{2015}\natexlab{}.
\newblock \bibinfo{title}{Expanding the Vocabulary of Multitouch Input using Magnetic Fingerprints}.
\newblock
\newblock
\showeprint[arxiv]{1501.03218}~[cs.HC]


\bibitem[Athavipach et~al\mbox{.}(2019)]%
        {athavipach2019wearable}
\bibfield{author}{\bibinfo{person}{Chanavit Athavipach}, \bibinfo{person}{Setha Pan-ngum}, {and} \bibinfo{person}{Pasin Israsena}.} \bibinfo{year}{2019}\natexlab{}.
\newblock \showarticletitle{A Wearable In-Ear EEG Device for Emotion Monitoring}.
\newblock \bibinfo{journal}{\emph{Sensors}} \bibinfo{volume}{19}, \bibinfo{number}{18} (\bibinfo{year}{2019}).
\newblock
\showISSN{1424-8220}
\urldef\tempurl%
\url{https://doi.org/10.3390/s19184014}
\showDOI{\tempurl}


\bibitem[Au and Tai(2010)]%
        {au2010multitouch}
\bibfield{author}{\bibinfo{person}{Oscar Kin-Chung Au} {and} \bibinfo{person}{Chiew-Lan Tai}.} \bibinfo{year}{2010}\natexlab{}.
\newblock \showarticletitle{Multitouch Finger Registration and Its Applications}. In \bibinfo{booktitle}{\emph{Proceedings of the 22nd Conference of the Computer-Human Interaction Special Interest Group of Australia on Computer-Human Interaction}} (Brisbane, Australia) \emph{(\bibinfo{series}{OZCHI '10})}. \bibinfo{publisher}{Association for Computing Machinery}, \bibinfo{address}{New York, NY, USA}, \bibinfo{pages}{41–48}.
\newblock
\showISBNx{9781450305020}
\urldef\tempurl%
\url{https://doi.org/10.1145/1952222.1952233}
\showDOI{\tempurl}


\bibitem[Ba~h et~al\mbox{.}(2008)]%
        {touchLook08Ba}
\bibfield{author}{\bibinfo{person}{Kenneth~Majlund Ba~h}, \bibinfo{person}{Mads~Gregers J\ae{}ger}, \bibinfo{person}{Mikael~B. Skov}, {and} \bibinfo{person}{Nils~Gram Thomassen}.} \bibinfo{year}{2008}\natexlab{}.
\newblock \showarticletitle{You can touch, but you can't look: interacting with in-vehicle systems}. In \bibinfo{booktitle}{\emph{Proceedings of the SIGCHI Conference on Human Factors in Computing Systems}} (Florence, Italy) \emph{(\bibinfo{series}{CHI '08})}. \bibinfo{publisher}{Association for Computing Machinery}, \bibinfo{address}{New York, NY, USA}, \bibinfo{pages}{1139–1148}.
\newblock
\showISBNx{9781605580111}
\urldef\tempurl%
\url{https://doi.org/10.1145/1357054.1357233}
\showDOI{\tempurl}


\bibitem[Benko et~al\mbox{.}(2009)]%
        {benko2009enhancing}
\bibfield{author}{\bibinfo{person}{Hrvoje Benko}, \bibinfo{person}{T.~Scott Saponas}, \bibinfo{person}{Dan Morris}, {and} \bibinfo{person}{Desney Tan}.} \bibinfo{year}{2009}\natexlab{}.
\newblock \showarticletitle{Enhancing Input on and above the Interactive Surface with Muscle Sensing}. In \bibinfo{booktitle}{\emph{Proceedings of the ACM International Conference on Interactive Tabletops and Surfaces}} (Banff, Alberta, Canada) \emph{(\bibinfo{series}{ITS '09})}. \bibinfo{publisher}{Association for Computing Machinery}, \bibinfo{address}{New York, NY, USA}, \bibinfo{pages}{93–100}.
\newblock
\showISBNx{9781605587332}
\urldef\tempurl%
\url{https://doi.org/10.1145/1731903.1731924}
\showDOI{\tempurl}


\bibitem[Brooke(1995)]%
        {brooke1996sus}
\bibfield{author}{\bibinfo{person}{John Brooke}.} \bibinfo{year}{1995}\natexlab{}.
\newblock \showarticletitle{SUS: A quick and dirty usability scale}.
\newblock \bibinfo{journal}{\emph{Usability Eval. Ind.}}  \bibinfo{volume}{189} (\bibinfo{date}{11} \bibinfo{year}{1995}).
\newblock


\bibitem[Brudy et~al\mbox{.}(2019)]%
        {CrossDeviceTaxonomyCHI19}
\bibfield{author}{\bibinfo{person}{Frederik Brudy}, \bibinfo{person}{Christian Holz}, \bibinfo{person}{Roman R\"{a}dle}, \bibinfo{person}{Chi-Jui Wu}, \bibinfo{person}{Steven Houben}, \bibinfo{person}{Clemens~Nylandsted Klokmose}, {and} \bibinfo{person}{Nicolai Marquardt}.} \bibinfo{year}{2019}\natexlab{}.
\newblock \showarticletitle{Cross-Device Taxonomy: Survey, Opportunities and Challenges of Interactions Spanning Across Multiple Devices}. In \bibinfo{booktitle}{\emph{Proceedings of the 2019 CHI Conference on Human Factors in Computing Systems}} (Glasgow, Scotland Uk) \emph{(\bibinfo{series}{CHI '19})}. \bibinfo{publisher}{Association for Computing Machinery}, \bibinfo{address}{New York, NY, USA}, \bibinfo{pages}{1–28}.
\newblock
\showISBNx{9781450359702}
\urldef\tempurl%
\url{https://doi.org/10.1145/3290605.3300792}
\showDOI{\tempurl}


\bibitem[Chen et~al\mbox{.}(2020)]%
        {listen2020chen}
\bibfield{author}{\bibinfo{person}{Huijie Chen}, \bibinfo{person}{Fan Li}, \bibinfo{person}{Wan Du}, \bibinfo{person}{Song Yang}, \bibinfo{person}{Matthew Conn}, {and} \bibinfo{person}{Yu Wang}.} \bibinfo{year}{2020}\natexlab{}.
\newblock \showarticletitle{Listen to Your Fingers: User Authentication Based on Geometry Biometrics of Touch Gesture}.
\newblock \bibinfo{journal}{\emph{Proc. ACM Interact. Mob. Wearable Ubiquitous Technol.}} \bibinfo{volume}{4}, \bibinfo{number}{3}, Article \bibinfo{articleno}{75} (\bibinfo{date}{sep} \bibinfo{year}{2020}), \bibinfo{numpages}{23}~pages.
\newblock
\urldef\tempurl%
\url{https://doi.org/10.1145/3411809}
\showDOI{\tempurl}


\bibitem[Chen et~al\mbox{.}(2013)]%
        {chen2013utrack}
\bibfield{author}{\bibinfo{person}{Ke-Yu Chen}, \bibinfo{person}{Kent Lyons}, \bibinfo{person}{Sean White}, {and} \bibinfo{person}{Shwetak Patel}.} \bibinfo{year}{2013}\natexlab{}.
\newblock \showarticletitle{UTrack: 3D Input Using Two Magnetic Sensors}. In \bibinfo{booktitle}{\emph{Proceedings of the 26th Annual ACM Symposium on User Interface Software and Technology}} (St. Andrews, Scotland, United Kingdom) \emph{(\bibinfo{series}{UIST '13})}. \bibinfo{publisher}{Association for Computing Machinery}, \bibinfo{address}{New York, NY, USA}, \bibinfo{pages}{237–244}.
\newblock
\showISBNx{9781450322683}
\urldef\tempurl%
\url{https://doi.org/10.1145/2501988.2502035}
\showDOI{\tempurl}


\bibitem[Chen et~al\mbox{.}(2016)]%
        {chen2016finexus}
\bibfield{author}{\bibinfo{person}{Ke-Yu Chen}, \bibinfo{person}{Shwetak~N. Patel}, {and} \bibinfo{person}{Sean Keller}.} \bibinfo{year}{2016}\natexlab{}.
\newblock \showarticletitle{Finexus: Tracking Precise Motions of Multiple Fingertips Using Magnetic Sensing}. In \bibinfo{booktitle}{\emph{Proceedings of the 2016 CHI Conference on Human Factors in Computing Systems}} (San Jose, California, USA) \emph{(\bibinfo{series}{CHI '16})}. \bibinfo{publisher}{Association for Computing Machinery}, \bibinfo{address}{New York, NY, USA}, \bibinfo{pages}{1504–1514}.
\newblock
\showISBNx{9781450333627}
\urldef\tempurl%
\url{https://doi.org/10.1145/2858036.2858125}
\showDOI{\tempurl}


\bibitem[Cheung and Girouard(2019)]%
        {MagRingMobileMHCI19}
\bibfield{author}{\bibinfo{person}{Victor Cheung} {and} \bibinfo{person}{Audrey Girouard}.} \bibinfo{year}{2019}\natexlab{}.
\newblock \showarticletitle{Tangible Around-Device Interaction Using Rotatory Gestures with a Magnetic Ring}. In \bibinfo{booktitle}{\emph{Proceedings of the 21st International Conference on Human-Computer Interaction with Mobile Devices and Services}} (Taipei, Taiwan) \emph{(\bibinfo{series}{MobileHCI '19})}. \bibinfo{publisher}{Association for Computing Machinery}, \bibinfo{address}{New York, NY, USA}, Article \bibinfo{articleno}{26}, \bibinfo{numpages}{8}~pages.
\newblock
\showISBNx{9781450368254}
\urldef\tempurl%
\url{https://doi.org/10.1145/3338286.3340137}
\showDOI{\tempurl}


\bibitem[Choi et~al\mbox{.}(2023)]%
        {choi2023earppg}
\bibfield{author}{\bibinfo{person}{Seokmin Choi}, \bibinfo{person}{Junghwan Yim}, \bibinfo{person}{Yincheng Jin}, \bibinfo{person}{Yang Gao}, \bibinfo{person}{Jiyang Li}, {and} \bibinfo{person}{Zhanpeng Jin}.} \bibinfo{year}{2023}\natexlab{}.
\newblock \showarticletitle{EarPPG: Securing Your Identity with Your Ears}. In \bibinfo{booktitle}{\emph{Proceedings of the 28th International Conference on Intelligent User Interfaces}} (Sydney, NSW, Australia) \emph{(\bibinfo{series}{IUI '23})}. \bibinfo{publisher}{Association for Computing Machinery}, \bibinfo{address}{New York, NY, USA}, \bibinfo{pages}{835–849}.
\newblock
\showISBNx{9798400701061}
\urldef\tempurl%
\url{https://doi.org/10.1145/3581641.3584070}
\showDOI{\tempurl}


\bibitem[Curran et~al\mbox{.}(2016)]%
        {curran2016passthoughts}
\bibfield{author}{\bibinfo{person}{Max~T. Curran}, \bibinfo{person}{Jong-kai Yang}, \bibinfo{person}{Nick Merrill}, {and} \bibinfo{person}{John Chuang}.} \bibinfo{year}{2016}\natexlab{}.
\newblock \showarticletitle{Passthoughts authentication with low cost EarEEG}. In \bibinfo{booktitle}{\emph{2016 38th Annual International Conference of the IEEE Engineering in Medicine and Biology Society (EMBC)}}. \bibinfo{pages}{1979--1982}.
\newblock
\urldef\tempurl%
\url{https://doi.org/10.1109/EMBC.2016.7591112}
\showDOI{\tempurl}


\bibitem[Dobbelstein et~al\mbox{.}(2017)]%
        {Dobbelstein17}
\bibfield{author}{\bibinfo{person}{David Dobbelstein}, \bibinfo{person}{Gabriel Haas}, {and} \bibinfo{person}{Enrico Rukzio}.} \bibinfo{year}{2017}\natexlab{}.
\newblock \showarticletitle{The Effects of Mobility, Encumbrance, and (Non-)Dominant Hand on Interaction with Smartwatches}. In \bibinfo{booktitle}{\emph{Proceedings of the 2017 ACM International Symposium on Wearable Computers}} (Maui, Hawaii) \emph{(\bibinfo{series}{ISWC '17})}. \bibinfo{publisher}{Association for Computing Machinery}, \bibinfo{address}{New York, NY, USA}, \bibinfo{pages}{90–93}.
\newblock
\showISBNx{9781450351881}
\urldef\tempurl%
\url{https://doi.org/10.1145/3123021.3123033}
\showDOI{\tempurl}


\bibitem[Easwara~Moorthy and Vu(2014)]%
        {easwara2014voice}
\bibfield{author}{\bibinfo{person}{Aarthi Easwara~Moorthy} {and} \bibinfo{person}{Kim-Phuong~L. Vu}.} \bibinfo{year}{2014}\natexlab{}.
\newblock \showarticletitle{Voice Activated Personal Assistant: Acceptability of Use in the Public Space}. In \bibinfo{booktitle}{\emph{Human Interface and the Management of Information. Information and Knowledge in Applications and Services}}, \bibfield{editor}{\bibinfo{person}{Sakae Yamamoto}} (Ed.). \bibinfo{publisher}{Springer International Publishing}, \bibinfo{address}{Cham}, \bibinfo{pages}{324--334}.
\newblock
\showISBNx{978-3-319-07863-2}


\bibitem[Efthymiou and Halvey(2016)]%
        {efthymiou2016evaluating}
\bibfield{author}{\bibinfo{person}{Christos Efthymiou} {and} \bibinfo{person}{Martin Halvey}.} \bibinfo{year}{2016}\natexlab{}.
\newblock \showarticletitle{Evaluating the Social Acceptability of Voice Based Smartwatch Search}. In \bibinfo{booktitle}{\emph{Information Retrieval Technology: 12th Asia Information Retrieval Societies Conference, AIRS 2016, Beijing, China, November 30 – December 2, 2016, Proceedings}} (Beijing, China). \bibinfo{publisher}{Springer-Verlag}, \bibinfo{address}{Berlin, Heidelberg}, \bibinfo{pages}{267–278}.
\newblock
\showISBNx{978-3-319-48050-3}
\urldef\tempurl%
\url{https://doi.org/10.1007/978-3-319-48051-0_20}
\showDOI{\tempurl}


\bibitem[Ewerling et~al\mbox{.}(2012)]%
        {Multitouch1}
\bibfield{author}{\bibinfo{person}{Philipp Ewerling}, \bibinfo{person}{Alexander Kulik}, {and} \bibinfo{person}{Bernd Froehlich}.} \bibinfo{year}{2012}\natexlab{}.
\newblock \showarticletitle{Finger and Hand Detection for Multi-Touch Interfaces Based on Maximally Stable Extremal Regions}. In \bibinfo{booktitle}{\emph{Proceedings of the 2012 ACM International Conference on Interactive Tabletops and Surfaces}} (Cambridge, Massachusetts, USA) \emph{(\bibinfo{series}{ITS '12})}. \bibinfo{publisher}{Association for Computing Machinery}, \bibinfo{address}{New York, NY, USA}, \bibinfo{pages}{173–182}.
\newblock
\showISBNx{9781450312097}
\urldef\tempurl%
\url{https://doi.org/10.1145/2396636.2396663}
\showDOI{\tempurl}


\bibitem[Feld and Plummer(2019)]%
        {FELD2019219}
\bibfield{author}{\bibinfo{person}{Jody~A. Feld} {and} \bibinfo{person}{Prudence Plummer}.} \bibinfo{year}{2019}\natexlab{}.
\newblock \showarticletitle{Visual scanning behavior during distracted walking in healthy young adults}.
\newblock \bibinfo{journal}{\emph{Gait \& Posture}}  \bibinfo{volume}{67} (\bibinfo{year}{2019}), \bibinfo{pages}{219--223}.
\newblock
\showISSN{0966-6362}
\urldef\tempurl%
\url{https://doi.org/10.1016/j.gaitpost.2018.10.017}
\showDOI{\tempurl}


\bibitem[Ferlini et~al\mbox{.}(2021)]%
        {Ferlini21}
\bibfield{author}{\bibinfo{person}{Andrea Ferlini}, \bibinfo{person}{Alessandro Montanari}, \bibinfo{person}{Andreas Grammenos}, \bibinfo{person}{Robert Harle}, {and} \bibinfo{person}{Cecilia Mascolo}.} \bibinfo{year}{2021}\natexlab{}.
\newblock \showarticletitle{Enabling In-Ear Magnetic Sensing: Automatic and User Transparent Magnetometer Calibration}. In \bibinfo{booktitle}{\emph{2021 IEEE International Conference on Pervasive Computing and Communications (PerCom)}}. \bibinfo{pages}{1--8}.
\newblock
\urldef\tempurl%
\url{https://doi.org/10.1109/PERCOM50583.2021.9439112}
\showDOI{\tempurl}


\bibitem[Gao et~al\mbox{.}(2019)]%
        {gao2019earecho}
\bibfield{author}{\bibinfo{person}{Yang Gao}, \bibinfo{person}{Wei Wang}, \bibinfo{person}{Vir~V. Phoha}, \bibinfo{person}{Wei Sun}, {and} \bibinfo{person}{Zhanpeng Jin}.} \bibinfo{year}{2019}\natexlab{}.
\newblock \showarticletitle{EarEcho: Using Ear Canal Echo for Wearable Authentication}.
\newblock \bibinfo{journal}{\emph{Proc. ACM Interact. Mob. Wearable Ubiquitous Technol.}} \bibinfo{volume}{3}, \bibinfo{number}{3}, Article \bibinfo{articleno}{81} (\bibinfo{date}{sep} \bibinfo{year}{2019}), \bibinfo{numpages}{24}~pages.
\newblock
\urldef\tempurl%
\url{https://doi.org/10.1145/3351239}
\showDOI{\tempurl}


\bibitem[Gao et~al\mbox{.}(2022)]%
        {9826109}
\bibfield{author}{\bibinfo{person}{Zhihui Gao}, \bibinfo{person}{Ang Li}, \bibinfo{person}{Dong Li}, \bibinfo{person}{Jialin Liu}, \bibinfo{person}{Jie Xiong}, \bibinfo{person}{Yu Wang}, \bibinfo{person}{Bing Li}, {and} \bibinfo{person}{Yiran Chen}.} \bibinfo{year}{2022}\natexlab{}.
\newblock \showarticletitle{MOM: Microphone based 3D Orientation Measurement}. In \bibinfo{booktitle}{\emph{2022 21st ACM/IEEE International Conference on Information Processing in Sensor Networks (IPSN)}}. \bibinfo{pages}{132--144}.
\newblock
\urldef\tempurl%
\url{https://doi.org/10.1109/IPSN54338.2022.00018}
\showDOI{\tempurl}


\bibitem[Gil et~al\mbox{.}(2017)]%
        {gil2017tritap}
\bibfield{author}{\bibinfo{person}{Hyunjae Gil}, \bibinfo{person}{DoYoung Lee}, \bibinfo{person}{Seunggyu Im}, {and} \bibinfo{person}{Ian Oakley}.} \bibinfo{year}{2017}\natexlab{}.
\newblock \showarticletitle{TriTap: Identifying Finger Touches on Smartwatches}. In \bibinfo{booktitle}{\emph{Proceedings of the 2017 CHI Conference on Human Factors in Computing Systems}} (Denver, Colorado, USA) \emph{(\bibinfo{series}{CHI '17})}. \bibinfo{publisher}{Association for Computing Machinery}, \bibinfo{address}{New York, NY, USA}, \bibinfo{pages}{3879–3890}.
\newblock
\showISBNx{9781450346559}
\urldef\tempurl%
\url{https://doi.org/10.1145/3025453.3025561}
\showDOI{\tempurl}


\bibitem[Goguey et~al\mbox{.}(2016a)]%
        {goguey2016performance}
\bibfield{author}{\bibinfo{person}{Alix Goguey}, \bibinfo{person}{Mathieu Nancel}, \bibinfo{person}{G\'{e}ry Casiez}, {and} \bibinfo{person}{Daniel Vogel}.} \bibinfo{year}{2016}\natexlab{a}.
\newblock \showarticletitle{The Performance and Preference of Different Fingers and Chords for Pointing, Dragging, and Object Transformation}. In \bibinfo{booktitle}{\emph{Proceedings of the 2016 CHI Conference on Human Factors in Computing Systems}} (San Jose, California, USA) \emph{(\bibinfo{series}{CHI '16})}. \bibinfo{publisher}{Association for Computing Machinery}, \bibinfo{address}{New York, NY, USA}, \bibinfo{pages}{4250–4261}.
\newblock
\showISBNx{9781450333627}
\urldef\tempurl%
\url{https://doi.org/10.1145/2858036.2858194}
\showDOI{\tempurl}


\bibitem[Goguey et~al\mbox{.}(2016b)]%
        {Goguey16}
\bibfield{author}{\bibinfo{person}{Alix Goguey}, \bibinfo{person}{Mathieu Nancel}, \bibinfo{person}{G\'{e}ry Casiez}, {and} \bibinfo{person}{Daniel Vogel}.} \bibinfo{year}{2016}\natexlab{b}.
\newblock \showarticletitle{The Performance and Preference of Different Fingers and Chords for Pointing, Dragging, and Object Transformation}. In \bibinfo{booktitle}{\emph{Proceedings of the 2016 CHI Conference on Human Factors in Computing Systems}} (San Jose, California, USA) \emph{(\bibinfo{series}{CHI '16})}. \bibinfo{publisher}{Association for Computing Machinery}, \bibinfo{address}{New York, NY, USA}, \bibinfo{pages}{4250–4261}.
\newblock
\showISBNx{9781450333627}
\urldef\tempurl%
\url{https://doi.org/10.1145/2858036.2858194}
\showDOI{\tempurl}


\bibitem[Gong et~al\mbox{.}(2021)]%
        {gong2021robust}
\bibfield{author}{\bibinfo{person}{Jian Gong}, \bibinfo{person}{Xinyu Zhang}, \bibinfo{person}{Yuanjun Huang}, \bibinfo{person}{Ju Ren}, {and} \bibinfo{person}{Yaoxue Zhang}.} \bibinfo{year}{2021}\natexlab{}.
\newblock \showarticletitle{Robust Inertial Motion Tracking through Deep Sensor Fusion across Smart Earbuds and Smartphone}.
\newblock \bibinfo{journal}{\emph{Proc. ACM Interact. Mob. Wearable Ubiquitous Technol.}} \bibinfo{volume}{5}, \bibinfo{number}{2}, Article \bibinfo{articleno}{62} (\bibinfo{date}{jun} \bibinfo{year}{2021}), \bibinfo{numpages}{26}~pages.
\newblock
\urldef\tempurl%
\url{https://doi.org/10.1145/3463517}
\showDOI{\tempurl}


\bibitem[Grier(2015)]%
        {Grier15}
\bibfield{author}{\bibinfo{person}{Rebecca~A. Grier}.} \bibinfo{year}{2015}\natexlab{}.
\newblock \showarticletitle{How High is High? A Meta-Analysis of NASA-TLX Global Workload Scores}.
\newblock \bibinfo{journal}{\emph{Proceedings of the Human Factors and Ergonomics Society Annual Meeting}} \bibinfo{volume}{59}, \bibinfo{number}{1} (\bibinfo{year}{2015}), \bibinfo{pages}{1727--1731}.
\newblock
\urldef\tempurl%
\url{https://doi.org/10.1177/1541931215591373}
\showDOI{\tempurl}
\showeprint{https://doi.org/10.1177/1541931215591373}


\bibitem[Gupta et~al\mbox{.}(2016)]%
        {gupta2016porous}
\bibfield{author}{\bibinfo{person}{Aakar Gupta}, \bibinfo{person}{Muhammed Anwar}, {and} \bibinfo{person}{Ravin Balakrishnan}.} \bibinfo{year}{2016}\natexlab{}.
\newblock \showarticletitle{Porous Interfaces for Small Screen Multitasking Using Finger Identification}. In \bibinfo{booktitle}{\emph{Proceedings of the 29th Annual Symposium on User Interface Software and Technology}} (Tokyo, Japan) \emph{(\bibinfo{series}{UIST '16})}. \bibinfo{publisher}{Association for Computing Machinery}, \bibinfo{address}{New York, NY, USA}, \bibinfo{pages}{145–156}.
\newblock
\showISBNx{9781450341899}
\urldef\tempurl%
\url{https://doi.org/10.1145/2984511.2984557}
\showDOI{\tempurl}


\bibitem[Gupta and Balakrishnan(2016)]%
        {gupta2016dualkey}
\bibfield{author}{\bibinfo{person}{Aakar Gupta} {and} \bibinfo{person}{Ravin Balakrishnan}.} \bibinfo{year}{2016}\natexlab{}.
\newblock \showarticletitle{DualKey: Miniature Screen Text Entry via Finger Identification}. In \bibinfo{booktitle}{\emph{Proceedings of the 2016 CHI Conference on Human Factors in Computing Systems}} (San Jose, California, USA) \emph{(\bibinfo{series}{CHI '16})}. \bibinfo{publisher}{Association for Computing Machinery}, \bibinfo{address}{New York, NY, USA}, \bibinfo{pages}{59–70}.
\newblock
\showISBNx{9781450333627}
\urldef\tempurl%
\url{https://doi.org/10.1145/2858036.2858052}
\showDOI{\tempurl}


\bibitem[Harrison and Hudson(2009)]%
        {harrison2009abracadabra}
\bibfield{author}{\bibinfo{person}{Chris Harrison} {and} \bibinfo{person}{Scott~E. Hudson}.} \bibinfo{year}{2009}\natexlab{}.
\newblock \showarticletitle{Abracadabra: Wireless, High-Precision, and Unpowered Finger Input for Very Small Mobile Devices}. In \bibinfo{booktitle}{\emph{Proceedings of the 22nd Annual ACM Symposium on User Interface Software and Technology}} (Victoria, BC, Canada) \emph{(\bibinfo{series}{UIST '09})}. \bibinfo{publisher}{Association for Computing Machinery}, \bibinfo{address}{New York, NY, USA}, \bibinfo{pages}{121–124}.
\newblock
\showISBNx{9781605587455}
\urldef\tempurl%
\url{https://doi.org/10.1145/1622176.1622199}
\showDOI{\tempurl}


\bibitem[Hessar et~al\mbox{.}(2016)]%
        {hessar2016enabling}
\bibfield{author}{\bibinfo{person}{Mehrdad Hessar}, \bibinfo{person}{Vikram Iyer}, {and} \bibinfo{person}{Shyamnath Gollakota}.} \bibinfo{year}{2016}\natexlab{}.
\newblock \showarticletitle{Enabling On-Body Transmissions with Commodity Devices}. In \bibinfo{booktitle}{\emph{Proceedings of the 2016 ACM International Joint Conference on Pervasive and Ubiquitous Computing}} (Heidelberg, Germany) \emph{(\bibinfo{series}{UbiComp '16})}. \bibinfo{publisher}{Association for Computing Machinery}, \bibinfo{address}{New York, NY, USA}, \bibinfo{pages}{1100–1111}.
\newblock
\showISBNx{9781450344616}
\urldef\tempurl%
\url{https://doi.org/10.1145/2971648.2971682}
\showDOI{\tempurl}


\bibitem[Jimenez-Olmedo et~al\mbox{.}(2019)]%
        {interdigitalWeb19}
\bibfield{author}{\bibinfo{person}{Jose Jimenez-Olmedo}, \bibinfo{person}{Alfonso Penichet-Tomás}, \bibinfo{person}{Manuel Ortega-Becerra}, \bibinfo{person}{Basilio Pueo}, {and} \bibinfo{person}{Jose Espina-Agullo}.} \bibinfo{year}{2019}\natexlab{}.
\newblock \showarticletitle{Relationships between anthropometric parameters and overarm throw in elite beach handball}.
\newblock \bibinfo{journal}{\emph{Human Movement}}  \bibinfo{volume}{20} (\bibinfo{date}{04} \bibinfo{year}{2019}), \bibinfo{pages}{16--24}.
\newblock
\urldef\tempurl%
\url{https://doi.org/10.5114/hm.2019.79394}
\showDOI{\tempurl}


\bibitem[Kadomura and Siio(2015)]%
        {kadomura2015magnail}
\bibfield{author}{\bibinfo{person}{Azusa Kadomura} {and} \bibinfo{person}{Itiro Siio}.} \bibinfo{year}{2015}\natexlab{}.
\newblock \showarticletitle{MagNail: User Interaction with Smart Device through Magnet Attached to Fingernail}. In \bibinfo{booktitle}{\emph{Adjunct Proceedings of the 2015 ACM International Joint Conference on Pervasive and Ubiquitous Computing and Proceedings of the 2015 ACM International Symposium on Wearable Computers}} (Osaka, Japan) \emph{(\bibinfo{series}{UbiComp/ISWC'15 Adjunct})}. \bibinfo{publisher}{Association for Computing Machinery}, \bibinfo{address}{New York, NY, USA}, \bibinfo{pages}{309–312}.
\newblock
\showISBNx{9781450335751}
\urldef\tempurl%
\url{https://doi.org/10.1145/2800835.2800859}
\showDOI{\tempurl}


\bibitem[Katayama et~al\mbox{.}(2019)]%
        {katayama2019situation}
\bibfield{author}{\bibinfo{person}{Shin Katayama}, \bibinfo{person}{Akhil Mathur}, \bibinfo{person}{Marc van~den Broeck}, \bibinfo{person}{Tadashi Okoshi}, \bibinfo{person}{Jin Nakazawa}, {and} \bibinfo{person}{Fahim Kawsar}.} \bibinfo{year}{2019}\natexlab{}.
\newblock \showarticletitle{Situation-Aware Emotion Regulation of Conversational Agents with Kinetic Earables}. In \bibinfo{booktitle}{\emph{2019 8th International Conference on Affective Computing and Intelligent Interaction (ACII)}}. \bibinfo{pages}{725--731}.
\newblock
\showISSN{2156-8111}
\urldef\tempurl%
\url{https://doi.org/10.1109/ACII.2019.8925449}
\showDOI{\tempurl}


\bibitem[Ketabdar et~al\mbox{.}(2010)]%
        {TowardsUsing10MHCI}
\bibfield{author}{\bibinfo{person}{Hamed Ketabdar}, \bibinfo{person}{Mehran Roshandel}, {and} \bibinfo{person}{Kamer~Ali Y\"{u}ksel}.} \bibinfo{year}{2010}\natexlab{}.
\newblock \showarticletitle{Towards using embedded magnetic field sensor for around mobile device 3D interaction}. In \bibinfo{booktitle}{\emph{Proceedings of the 12th International Conference on Human Computer Interaction with Mobile Devices and Services}} (Lisbon, Portugal) \emph{(\bibinfo{series}{MobileHCI '10})}. \bibinfo{publisher}{Association for Computing Machinery}, \bibinfo{address}{New York, NY, USA}, \bibinfo{pages}{153–156}.
\newblock
\showISBNx{9781605588353}
\urldef\tempurl%
\url{https://doi.org/10.1145/1851600.1851626}
\showDOI{\tempurl}


\bibitem[Kikuchi et~al\mbox{.}(2017)]%
        {kikuchi2017eartouch}
\bibfield{author}{\bibinfo{person}{Takashi Kikuchi}, \bibinfo{person}{Yuta Sugiura}, \bibinfo{person}{Katsutoshi Masai}, \bibinfo{person}{Maki Sugimoto}, {and} \bibinfo{person}{Bruce~H. Thomas}.} \bibinfo{year}{2017}\natexlab{}.
\newblock \showarticletitle{EarTouch: Turning the Ear into an Input Surface}. In \bibinfo{booktitle}{\emph{Proceedings of the 19th International Conference on Human-Computer Interaction with Mobile Devices and Services}} (Vienna, Austria) \emph{(\bibinfo{series}{MobileHCI '17})}. \bibinfo{publisher}{Association for Computing Machinery}, \bibinfo{address}{New York, NY, USA}, Article \bibinfo{articleno}{27}, \bibinfo{numpages}{6}~pages.
\newblock
\showISBNx{9781450350754}
\urldef\tempurl%
\url{https://doi.org/10.1145/3098279.3098538}
\showDOI{\tempurl}


\bibitem[Kim and Oakley(2022)]%
        {kim2022sonarid}
\bibfield{author}{\bibinfo{person}{Jiwan Kim} {and} \bibinfo{person}{Ian Oakley}.} \bibinfo{year}{2022}\natexlab{}.
\newblock \showarticletitle{SonarID: Using Sonar to Identify Fingers on a Smartwatch}. In \bibinfo{booktitle}{\emph{Proceedings of the 2022 CHI Conference on Human Factors in Computing Systems}} (New Orleans, LA, USA) \emph{(\bibinfo{series}{CHI '22})}. \bibinfo{publisher}{Association for Computing Machinery}, \bibinfo{address}{New York, NY, USA}, Article \bibinfo{articleno}{287}, \bibinfo{numpages}{10}~pages.
\newblock
\showISBNx{9781450391573}
\urldef\tempurl%
\url{https://doi.org/10.1145/3491102.3501935}
\showDOI{\tempurl}


\bibitem[Lafreniere et~al\mbox{.}(2016)]%
        {Spatial2016Benjamin}
\bibfield{author}{\bibinfo{person}{Benjamin Lafreniere}, \bibinfo{person}{Carl Gutwin}, \bibinfo{person}{Andy Cockburn}, {and} \bibinfo{person}{Tovi Grossman}.} \bibinfo{year}{2016}\natexlab{}.
\newblock \showarticletitle{Faster Command Selection on Touchscreen Watches}. In \bibinfo{booktitle}{\emph{Proceedings of the 2016 CHI Conference on Human Factors in Computing Systems}} (San Jose, California, USA) \emph{(\bibinfo{series}{CHI '16})}. \bibinfo{publisher}{Association for Computing Machinery}, \bibinfo{address}{New York, NY, USA}, \bibinfo{pages}{4663–4674}.
\newblock
\showISBNx{9781450333627}
\urldef\tempurl%
\url{https://doi.org/10.1145/2858036.2858166}
\showDOI{\tempurl}


\bibitem[Le et~al\mbox{.}(2019)]%
        {le2019investigating}
\bibfield{author}{\bibinfo{person}{Huy~Viet Le}, \bibinfo{person}{Sven Mayer}, {and} \bibinfo{person}{Niels Henze}.} \bibinfo{year}{2019}\natexlab{}.
\newblock \showarticletitle{Investigating the Feasibility of Finger Identification on Capacitive Touchscreens Using Deep Learning}. In \bibinfo{booktitle}{\emph{Proceedings of the 24th International Conference on Intelligent User Interfaces}} (Marina del Ray, California) \emph{(\bibinfo{series}{IUI '19})}. \bibinfo{publisher}{Association for Computing Machinery}, \bibinfo{address}{New York, NY, USA}, \bibinfo{pages}{637–649}.
\newblock
\showISBNx{9781450362726}
\urldef\tempurl%
\url{https://doi.org/10.1145/3301275.3302295}
\showDOI{\tempurl}


\bibitem[Lee et~al\mbox{.}(2021)]%
        {FingerText21Lee}
\bibfield{author}{\bibinfo{person}{DoYoung Lee}, \bibinfo{person}{Jiwan Kim}, {and} \bibinfo{person}{Ian Oakley}.} \bibinfo{year}{2021}\natexlab{}.
\newblock \showarticletitle{FingerText: Exploring and Optimizing Performance for Wearable, Mobile and One-Handed Typing}. In \bibinfo{booktitle}{\emph{Proceedings of the 2021 CHI Conference on Human Factors in Computing Systems}} (<conf-loc>, <city>Yokohama</city>, <country>Japan</country>, </conf-loc>) \emph{(\bibinfo{series}{CHI '21})}. \bibinfo{publisher}{Association for Computing Machinery}, \bibinfo{address}{New York, NY, USA}, Article \bibinfo{articleno}{283}, \bibinfo{numpages}{15}~pages.
\newblock
\showISBNx{9781450380966}
\urldef\tempurl%
\url{https://doi.org/10.1145/3411764.3445106}
\showDOI{\tempurl}


\bibitem[Lee et~al\mbox{.}(2018)]%
        {Lee2018SocialAcc}
\bibfield{author}{\bibinfo{person}{DoYoung Lee}, \bibinfo{person}{Youryang Lee}, \bibinfo{person}{Yonghwan Shin}, {and} \bibinfo{person}{Ian Oakley}.} \bibinfo{year}{2018}\natexlab{}.
\newblock \showarticletitle{Designing Socially Acceptable Hand-to-Face Input}. In \bibinfo{booktitle}{\emph{Proceedings of the 31st Annual ACM Symposium on User Interface Software and Technology}} (Berlin, Germany) \emph{(\bibinfo{series}{UIST '18})}. \bibinfo{publisher}{Association for Computing Machinery}, \bibinfo{address}{New York, NY, USA}, \bibinfo{pages}{711–723}.
\newblock
\showISBNx{9781450359481}
\urldef\tempurl%
\url{https://doi.org/10.1145/3242587.3242642}
\showDOI{\tempurl}


\bibitem[Lewis and Sauro(2018)]%
        {lewis2018item}
\bibfield{author}{\bibinfo{person}{James~R. Lewis} {and} \bibinfo{person}{Jeff Sauro}.} \bibinfo{year}{2018}\natexlab{}.
\newblock \showarticletitle{Item benchmarks for the system usability scale}.
\newblock \bibinfo{journal}{\emph{J. Usability Studies}} \bibinfo{volume}{13}, \bibinfo{number}{3} (\bibinfo{date}{May} \bibinfo{year}{2018}), \bibinfo{pages}{158–167}.
\newblock


\bibitem[Li et~al\mbox{.}(2008)]%
        {BlindSight}
\bibfield{author}{\bibinfo{person}{Kevin~A. Li}, \bibinfo{person}{Patrick Baudisch}, {and} \bibinfo{person}{Ken Hinckley}.} \bibinfo{year}{2008}\natexlab{}.
\newblock \showarticletitle{Blindsight: Eyes-Free Access to Mobile Phones}. In \bibinfo{booktitle}{\emph{Proceedings of the SIGCHI Conference on Human Factors in Computing Systems}} (Florence, Italy) \emph{(\bibinfo{series}{CHI '08})}. \bibinfo{publisher}{Association for Computing Machinery}, \bibinfo{address}{New York, NY, USA}, \bibinfo{pages}{1389–1398}.
\newblock
\showISBNx{9781605580111}
\urldef\tempurl%
\url{https://doi.org/10.1145/1357054.1357273}
\showDOI{\tempurl}


\bibitem[Liang et~al\mbox{.}(2013)]%
        {liang2013gaussbits}
\bibfield{author}{\bibinfo{person}{Rong-Hao Liang}, \bibinfo{person}{Kai-Yin Cheng}, \bibinfo{person}{Liwei Chan}, \bibinfo{person}{Chuan-Xhyuan Peng}, \bibinfo{person}{Mike~Y. Chen}, \bibinfo{person}{Rung-Huei Liang}, \bibinfo{person}{De-Nian Yang}, {and} \bibinfo{person}{Bing-Yu Chen}.} \bibinfo{year}{2013}\natexlab{}.
\newblock \showarticletitle{GaussBits: Magnetic Tangible Bits for Portable and Occlusion-Free near-Surface Interactions}. In \bibinfo{booktitle}{\emph{Proceedings of the SIGCHI Conference on Human Factors in Computing Systems}} (Paris, France) \emph{(\bibinfo{series}{CHI '13})}. \bibinfo{publisher}{Association for Computing Machinery}, \bibinfo{address}{New York, NY, USA}, \bibinfo{pages}{1391–1400}.
\newblock
\showISBNx{9781450318990}
\urldef\tempurl%
\url{https://doi.org/10.1145/2470654.2466185}
\showDOI{\tempurl}


\bibitem[Lim and Feria(2012)]%
        {Walking12Lim}
\bibfield{author}{\bibinfo{person}{Ji~Jung Lim} {and} \bibinfo{person}{Cary Feria}.} \bibinfo{year}{2012}\natexlab{}.
\newblock \showarticletitle{Visual Search on a Mobile Device While Walking}. In \bibinfo{booktitle}{\emph{Proceedings of the 14th International Conference on Human-Computer Interaction with Mobile Devices and Services}} (San Francisco, California, USA) \emph{(\bibinfo{series}{MobileHCI '12})}. \bibinfo{publisher}{Association for Computing Machinery}, \bibinfo{address}{New York, NY, USA}, \bibinfo{pages}{295–304}.
\newblock
\showISBNx{9781450311052}
\urldef\tempurl%
\url{https://doi.org/10.1145/2371574.2371618}
\showDOI{\tempurl}


\bibitem[Lissermann et~al\mbox{.}(2014)]%
        {lissermann2014earput}
\bibfield{author}{\bibinfo{person}{Roman Lissermann}, \bibinfo{person}{Jochen Huber}, \bibinfo{person}{Aristotelis Hadjakos}, \bibinfo{person}{Suranga Nanayakkara}, {and} \bibinfo{person}{Max M\"{u}hlh\"{a}user}.} \bibinfo{year}{2014}\natexlab{}.
\newblock \showarticletitle{EarPut: Augmenting Ear-Worn Devices for Ear-Based Interaction}. In \bibinfo{booktitle}{\emph{Proceedings of the 26th Australian Computer-Human Interaction Conference on Designing Futures: The Future of Design}} (Sydney, New South Wales, Australia) \emph{(\bibinfo{series}{OzCHI '14})}. \bibinfo{publisher}{Association for Computing Machinery}, \bibinfo{address}{New York, NY, USA}, \bibinfo{pages}{300–307}.
\newblock
\showISBNx{9781450306539}
\urldef\tempurl%
\url{https://doi.org/10.1145/2686612.2686655}
\showDOI{\tempurl}


\bibitem[Makinistian et~al\mbox{.}(2022)]%
        {MAKINISTIAN2022113907}
\bibfield{author}{\bibinfo{person}{L. Makinistian}, \bibinfo{person}{L. Zastko}, \bibinfo{person}{A. Tvarožná}, \bibinfo{person}{L.E. Días}, {and} \bibinfo{person}{I. Belyaev}.} \bibinfo{year}{2022}\natexlab{}.
\newblock \showarticletitle{Static magnetic fields from earphones: Detailed measurements plus some open questions}.
\newblock \bibinfo{journal}{\emph{Environmental Research}}  \bibinfo{volume}{214} (\bibinfo{year}{2022}), \bibinfo{pages}{113907}.
\newblock
\showISSN{0013-9351}
\urldef\tempurl%
\url{https://doi.org/10.1016/j.envres.2022.113907}
\showDOI{\tempurl}


\bibitem[Martin and Voix(2018)]%
        {martin2017ear}
\bibfield{author}{\bibinfo{person}{Alexis Martin} {and} \bibinfo{person}{Jérémie Voix}.} \bibinfo{year}{2018}\natexlab{}.
\newblock \showarticletitle{In-Ear Audio Wearable: Measurement of Heart and Breathing Rates for Health and Safety Monitoring}.
\newblock \bibinfo{journal}{\emph{IEEE Transactions on Biomedical Engineering}} \bibinfo{volume}{65}, \bibinfo{number}{6} (\bibinfo{year}{2018}), \bibinfo{pages}{1256--1263}.
\newblock
\urldef\tempurl%
\url{https://doi.org/10.1109/TBME.2017.2720463}
\showDOI{\tempurl}


\bibitem[Masson et~al\mbox{.}(2017)]%
        {masson2017whichfingers}
\bibfield{author}{\bibinfo{person}{Damien Masson}, \bibinfo{person}{Alix Goguey}, \bibinfo{person}{Sylvain Malacria}, {and} \bibinfo{person}{G\'{e}ry Casiez}.} \bibinfo{year}{2017}\natexlab{}.
\newblock \showarticletitle{WhichFingers: Identifying Fingers on Touch Surfaces and Keyboards Using Vibration Sensors}. In \bibinfo{booktitle}{\emph{Proceedings of the 30th Annual ACM Symposium on User Interface Software and Technology}} (Qu\'{e}bec City, QC, Canada) \emph{(\bibinfo{series}{UIST '17})}. \bibinfo{publisher}{Association for Computing Machinery}, \bibinfo{address}{New York, NY, USA}, \bibinfo{pages}{41–48}.
\newblock
\showISBNx{9781450349819}
\urldef\tempurl%
\url{https://doi.org/10.1145/3126594.3126619}
\showDOI{\tempurl}


\bibitem[McIntosh et~al\mbox{.}(2019)]%
        {mcintosh2019magnetips}
\bibfield{author}{\bibinfo{person}{Jess McIntosh}, \bibinfo{person}{Paul Strohmeier}, \bibinfo{person}{Jarrod Knibbe}, \bibinfo{person}{Sebastian Boring}, {and} \bibinfo{person}{Kasper Hornb\ae{}k}.} \bibinfo{year}{2019}\natexlab{}.
\newblock \showarticletitle{Magnetips: Combining Fingertip Tracking and Haptic Feedback for Around-Device Interaction}. In \bibinfo{booktitle}{\emph{Proceedings of the 2019 CHI Conference on Human Factors in Computing Systems}} (Glasgow, Scotland Uk) \emph{(\bibinfo{series}{CHI '19})}. \bibinfo{publisher}{Association for Computing Machinery}, \bibinfo{address}{New York, NY, USA}, \bibinfo{pages}{1–12}.
\newblock
\showISBNx{9781450359702}
\urldef\tempurl%
\url{https://doi.org/10.1145/3290605.3300638}
\showDOI{\tempurl}


\bibitem[Metzger et~al\mbox{.}(2004)]%
        {metzger2004freedigiter}
\bibfield{author}{\bibinfo{person}{Christian Metzger}, \bibinfo{person}{Matt Anderson}, {and} \bibinfo{person}{Thad Starner}.} \bibinfo{year}{2004}\natexlab{}.
\newblock \showarticletitle{FreeDigiter: A Contact-Free Device for Gesture Control}. In \bibinfo{booktitle}{\emph{Proceedings of the Eighth International Symposium on Wearable Computers}} \emph{(\bibinfo{series}{ISWC '04})}. \bibinfo{publisher}{IEEE Computer Society}, \bibinfo{address}{USA}, \bibinfo{pages}{18–21}.
\newblock
\showISBNx{076952186X}
\urldef\tempurl%
\url{https://doi.org/10.1109/ISWC.2004.23}
\showDOI{\tempurl}


\bibitem[Montanari et~al\mbox{.}(2024)]%
        {montanari2024omnibudssensoryearableplatform}
\bibfield{author}{\bibinfo{person}{Alessandro Montanari}, \bibinfo{person}{Ashok Thangarajan}, \bibinfo{person}{Khaldoon Al-Naimi}, \bibinfo{person}{Andrea Ferlini}, \bibinfo{person}{Yang Liu}, \bibinfo{person}{Ananta~Narayanan Balaji}, {and} \bibinfo{person}{Fahim Kawsar}.} \bibinfo{year}{2024}\natexlab{}.
\newblock \bibinfo{title}{OmniBuds: A Sensory Earable Platform for Advanced Bio-Sensing and On-Device Machine Learning}.
\newblock
\newblock
\showeprint[arxiv]{2410.04775}~[cs.ET]
\urldef\tempurl%
\url{https://arxiv.org/abs/2410.04775}
\showURL{%
\tempurl}


\bibitem[Murugappan et~al\mbox{.}(2012)]%
        {Multitouch2}
\bibfield{author}{\bibinfo{person}{Sundar Murugappan}, \bibinfo{person}{Vinayak}, \bibinfo{person}{Niklas Elmqvist}, {and} \bibinfo{person}{Karthik Ramani}.} \bibinfo{year}{2012}\natexlab{}.
\newblock \showarticletitle{Extended Multitouch: Recovering Touch Posture and Differentiating Users Using a Depth Camera}. In \bibinfo{booktitle}{\emph{Proceedings of the 25th Annual ACM Symposium on User Interface Software and Technology}} (Cambridge, Massachusetts, USA) \emph{(\bibinfo{series}{UIST '12})}. \bibinfo{publisher}{Association for Computing Machinery}, \bibinfo{address}{New York, NY, USA}, \bibinfo{pages}{487–496}.
\newblock
\showISBNx{9781450315807}
\urldef\tempurl%
\url{https://doi.org/10.1145/2380116.2380177}
\showDOI{\tempurl}


\bibitem[Namnakani et~al\mbox{.}(2023)]%
        {Namnakani23}
\bibfield{author}{\bibinfo{person}{Omar Namnakani}, \bibinfo{person}{Yasmeen Abdrabou}, \bibinfo{person}{Jonathan Grizou}, \bibinfo{person}{Augusto Esteves}, {and} \bibinfo{person}{Mohamed Khamis}.} \bibinfo{year}{2023}\natexlab{}.
\newblock \showarticletitle{Comparing Dwell Time, Pursuits and Gaze Gestures for Gaze Interaction on Handheld Mobile Devices}. In \bibinfo{booktitle}{\emph{Proceedings of the 2023 CHI Conference on Human Factors in Computing Systems}} (<conf-loc>, <city>Hamburg</city>, <country>Germany</country>, </conf-loc>) \emph{(\bibinfo{series}{CHI '23})}. \bibinfo{publisher}{Association for Computing Machinery}, \bibinfo{address}{New York, NY, USA}, Article \bibinfo{articleno}{258}, \bibinfo{numpages}{17}~pages.
\newblock
\showISBNx{9781450394215}
\urldef\tempurl%
\url{https://doi.org/10.1145/3544548.3580871}
\showDOI{\tempurl}


\bibitem[Oakley et~al\mbox{.}(2015)]%
        {Beats2015Ian}
\bibfield{author}{\bibinfo{person}{Ian Oakley}, \bibinfo{person}{DoYoung Lee}, \bibinfo{person}{MD.~Rasel Islam}, {and} \bibinfo{person}{Augusto Esteves}.} \bibinfo{year}{2015}\natexlab{}.
\newblock \showarticletitle{Beats: Tapping Gestures for Smart Watches}. In \bibinfo{booktitle}{\emph{Proceedings of the 33rd Annual ACM Conference on Human Factors in Computing Systems}} (Seoul, Republic of Korea) \emph{(\bibinfo{series}{CHI '15})}. \bibinfo{publisher}{Association for Computing Machinery}, \bibinfo{address}{New York, NY, USA}, \bibinfo{pages}{1237–1246}.
\newblock
\showISBNx{9781450331456}
\urldef\tempurl%
\url{https://doi.org/10.1145/2702123.2702226}
\showDOI{\tempurl}


\bibitem[Oney et~al\mbox{.}(2013)]%
        {zoomboard13stephen}
\bibfield{author}{\bibinfo{person}{Stephen Oney}, \bibinfo{person}{Chris Harrison}, \bibinfo{person}{Amy Ogan}, {and} \bibinfo{person}{Jason Wiese}.} \bibinfo{year}{2013}\natexlab{}.
\newblock \showarticletitle{ZoomBoard: a diminutive qwerty soft keyboard using iterative zooming for ultra-small devices}. In \bibinfo{booktitle}{\emph{Proceedings of the SIGCHI Conference on Human Factors in Computing Systems}} (Paris, France) \emph{(\bibinfo{series}{CHI '13})}. \bibinfo{publisher}{Association for Computing Machinery}, \bibinfo{address}{New York, NY, USA}, \bibinfo{pages}{2799–2802}.
\newblock
\showISBNx{9781450318990}
\urldef\tempurl%
\url{https://doi.org/10.1145/2470654.2481387}
\showDOI{\tempurl}


\bibitem[Park et~al\mbox{.}(2020)]%
        {geunwoo2020magtouch}
\bibfield{author}{\bibinfo{person}{Keunwoo Park}, \bibinfo{person}{Daehwa Kim}, \bibinfo{person}{Seongkook Heo}, {and} \bibinfo{person}{Geehyuk Lee}.} \bibinfo{year}{2020}\natexlab{}.
\newblock \showarticletitle{MagTouch: Robust Finger Identification for a Smartwatch Using a Magnet Ring and a Built-in Magnetometer}. In \bibinfo{booktitle}{\emph{Proceedings of the 2020 CHI Conference on Human Factors in Computing Systems}} (Honolulu, HI, USA) \emph{(\bibinfo{series}{CHI '20})}. \bibinfo{publisher}{Association for Computing Machinery}, \bibinfo{address}{New York, NY, USA}, \bibinfo{pages}{1–13}.
\newblock
\showISBNx{9781450367080}
\urldef\tempurl%
\url{https://doi.org/10.1145/3313831.3376234}
\showDOI{\tempurl}


\bibitem[Pirhonen et~al\mbox{.}(2002)]%
        {gestural02prhonen}
\bibfield{author}{\bibinfo{person}{Antti Pirhonen}, \bibinfo{person}{Stephen Brewster}, {and} \bibinfo{person}{Christopher Holguin}.} \bibinfo{year}{2002}\natexlab{}.
\newblock \showarticletitle{Gestural and audio metaphors as a means of control for mobile devices}. In \bibinfo{booktitle}{\emph{Proceedings of the SIGCHI Conference on Human Factors in Computing Systems}} (Minneapolis, Minnesota, USA) \emph{(\bibinfo{series}{CHI '02})}. \bibinfo{publisher}{Association for Computing Machinery}, \bibinfo{address}{New York, NY, USA}, \bibinfo{pages}{291–298}.
\newblock
\showISBNx{1581134533}
\urldef\tempurl%
\url{https://doi.org/10.1145/503376.503428}
\showDOI{\tempurl}


\bibitem[Poh et~al\mbox{.}(2009)]%
        {poh2009heartphones}
\bibfield{author}{\bibinfo{person}{Ming-Zher Poh}, \bibinfo{person}{Kyunghee Kim}, \bibinfo{person}{Andrew~D. Goessling}, \bibinfo{person}{Nicholas~C. Swenson}, {and} \bibinfo{person}{Rosalind~W. Picard}.} \bibinfo{year}{2009}\natexlab{}.
\newblock \showarticletitle{Heartphones: Sensor Earphones and Mobile Application for Non-obtrusive Health Monitoring}. In \bibinfo{booktitle}{\emph{Proceedings of the 2009 International Symposium on Wearable Computers}} \emph{(\bibinfo{series}{ISWC '09})}. \bibinfo{publisher}{IEEE Computer Society}, \bibinfo{address}{USA}, \bibinfo{pages}{153–154}.
\newblock
\showISBNx{9780769537795}
\urldef\tempurl%
\url{https://doi.org/10.1109/ISWC.2009.35}
\showDOI{\tempurl}


\bibitem[Rahman et~al\mbox{.}(2022)]%
        {rahman2022breathebuddy}
\bibfield{author}{\bibinfo{person}{Md~Mahbubur Rahman}, \bibinfo{person}{Tousif Ahmed}, \bibinfo{person}{Mohsin~Yusuf Ahmed}, \bibinfo{person}{Minh Dinh}, \bibinfo{person}{Ebrahim Nemati}, \bibinfo{person}{Jilong Kuang}, {and} \bibinfo{person}{Jun~Alex Gao}.} \bibinfo{year}{2022}\natexlab{}.
\newblock \showarticletitle{BreatheBuddy: Tracking Real-Time Breathing Exercises for Automated Biofeedback Using Commodity Earbuds}.
\newblock \bibinfo{journal}{\emph{Proc. ACM Hum.-Comput. Interact.}} \bibinfo{volume}{6}, \bibinfo{number}{MHCI}, Article \bibinfo{articleno}{213} (\bibinfo{date}{sep} \bibinfo{year}{2022}), \bibinfo{numpages}{18}~pages.
\newblock
\urldef\tempurl%
\url{https://doi.org/10.1145/3546748}
\showDOI{\tempurl}


\bibitem[Rateau et~al\mbox{.}(2022)]%
        {rateau2022leveraging}
\bibfield{author}{\bibinfo{person}{Hanae Rateau}, \bibinfo{person}{Edward Lank}, {and} \bibinfo{person}{Zhe Liu}.} \bibinfo{year}{2022}\natexlab{}.
\newblock \showarticletitle{Leveraging Smartwatch and Earbuds Gesture Capture to Support Wearable Interaction}.
\newblock \bibinfo{journal}{\emph{Proc. ACM Hum.-Comput. Interact.}} \bibinfo{volume}{6}, \bibinfo{number}{ISS}, Article \bibinfo{articleno}{557} (\bibinfo{date}{nov} \bibinfo{year}{2022}), \bibinfo{numpages}{20}~pages.
\newblock
\urldef\tempurl%
\url{https://doi.org/10.1145/3567710}
\showDOI{\tempurl}


\bibitem[Roy et~al\mbox{.}(2015)]%
        {roy2015glass+}
\bibfield{author}{\bibinfo{person}{Quentin Roy}, \bibinfo{person}{Yves Guiard}, \bibinfo{person}{Gilles Bailly}, \bibinfo{person}{{\'E}ric Lecolinet}, {and} \bibinfo{person}{Olivier Rioul}.} \bibinfo{year}{2015}\natexlab{}.
\newblock \showarticletitle{Glass+Skin: An Empirical Evaluation of the Added Value of Finger Identification to Basic Single-Touch Interaction on Touch Screens}. In \bibinfo{booktitle}{\emph{Human-Computer Interaction -- INTERACT 2015}}, \bibfield{editor}{\bibinfo{person}{Julio Abascal}, \bibinfo{person}{Simone Barbosa}, \bibinfo{person}{Mirko Fetter}, \bibinfo{person}{Tom Gross}, \bibinfo{person}{Philippe Palanque}, {and} \bibinfo{person}{Marco Winckler}} (Eds.). \bibinfo{publisher}{Springer International Publishing}, \bibinfo{address}{Cham}, \bibinfo{pages}{55--71}.
\newblock
\showISBNx{978-3-319-22723-8}


\bibitem[Siek et~al\mbox{.}(2005)]%
        {siek2005fat}
\bibfield{author}{\bibinfo{person}{Katie~A. Siek}, \bibinfo{person}{Yvonne Rogers}, {and} \bibinfo{person}{Kay~H. Connelly}.} \bibinfo{year}{2005}\natexlab{}.
\newblock \showarticletitle{Fat Finger Worries: How Older and Younger Users Physically Interact with PDAs}. In \bibinfo{booktitle}{\emph{Human-Computer Interaction - INTERACT 2005}}, \bibfield{editor}{\bibinfo{person}{Maria~Francesca Costabile} {and} \bibinfo{person}{Fabio Patern{\`o}}} (Eds.). \bibinfo{publisher}{Springer Berlin Heidelberg}, \bibinfo{address}{Berlin, Heidelberg}, \bibinfo{pages}{267--280}.
\newblock
\showISBNx{978-3-540-31722-7}


\bibitem[Singh et~al\mbox{.}(2018)]%
        {dswime18singh}
\bibfield{author}{\bibinfo{person}{Gaganpreet Singh}, \bibinfo{person}{William Delamare}, {and} \bibinfo{person}{Pourang Irani}.} \bibinfo{year}{2018}\natexlab{}.
\newblock \showarticletitle{D-SWIME: A Design Space for Smartwatch Interaction Techniques Supporting Mobility and Encumbrance}. In \bibinfo{booktitle}{\emph{Proceedings of the 2018 CHI Conference on Human Factors in Computing Systems}} (<conf-loc>, <city>Montreal QC</city>, <country>Canada</country>, </conf-loc>) \emph{(\bibinfo{series}{CHI '18})}. \bibinfo{publisher}{Association for Computing Machinery}, \bibinfo{address}{New York, NY, USA}, \bibinfo{pages}{1–13}.
\newblock
\showISBNx{9781450356206}
\urldef\tempurl%
\url{https://doi.org/10.1145/3173574.3174208}
\showDOI{\tempurl}


\bibitem[Sridhar et~al\mbox{.}(2017)]%
        {watchSense17Sridhar}
\bibfield{author}{\bibinfo{person}{Srinath Sridhar}, \bibinfo{person}{Anders Markussen}, \bibinfo{person}{Antti Oulasvirta}, \bibinfo{person}{Christian Theobalt}, {and} \bibinfo{person}{Sebastian Boring}.} \bibinfo{year}{2017}\natexlab{}.
\newblock \showarticletitle{WatchSense: On- and Above-Skin Input Sensing through a Wearable Depth Sensor}. In \bibinfo{booktitle}{\emph{Proceedings of the 2017 CHI Conference on Human Factors in Computing Systems}} (Denver, Colorado, USA) \emph{(\bibinfo{series}{CHI '17})}. \bibinfo{publisher}{Association for Computing Machinery}, \bibinfo{address}{New York, NY, USA}, \bibinfo{pages}{3891–3902}.
\newblock
\showISBNx{9781450346559}
\urldef\tempurl%
\url{https://doi.org/10.1145/3025453.3026005}
\showDOI{\tempurl}


\bibitem[Tsai et~al\mbox{.}(2016)]%
        {ThumbRingMHCI16}
\bibfield{author}{\bibinfo{person}{Hsin-Ruey Tsai}, \bibinfo{person}{Cheng-Yuan Wu}, \bibinfo{person}{Lee-Ting Huang}, {and} \bibinfo{person}{Yi-Ping Hung}.} \bibinfo{year}{2016}\natexlab{}.
\newblock \showarticletitle{ThumbRing: private interactions using one-handed thumb motion input on finger segments}. In \bibinfo{booktitle}{\emph{Proceedings of the 18th International Conference on Human-Computer Interaction with Mobile Devices and Services Adjunct}} (Florence, Italy) \emph{(\bibinfo{series}{MobileHCI '16})}. \bibinfo{publisher}{Association for Computing Machinery}, \bibinfo{address}{New York, NY, USA}, \bibinfo{pages}{791–798}.
\newblock
\showISBNx{9781450344135}
\urldef\tempurl%
\url{https://doi.org/10.1145/2957265.2961859}
\showDOI{\tempurl}


\bibitem[Wang et~al\mbox{.}(2021)]%
        {wang2021eardynamic}
\bibfield{author}{\bibinfo{person}{Zi Wang}, \bibinfo{person}{Sheng Tan}, \bibinfo{person}{Linghan Zhang}, \bibinfo{person}{Yili Ren}, \bibinfo{person}{Zhi Wang}, {and} \bibinfo{person}{Jie Yang}.} \bibinfo{year}{2021}\natexlab{}.
\newblock \showarticletitle{EarDynamic: An Ear Canal Deformation Based Continuous User Authentication Using In-Ear Wearables}.
\newblock \bibinfo{journal}{\emph{Proc. ACM Interact. Mob. Wearable Ubiquitous Technol.}} \bibinfo{volume}{5}, \bibinfo{number}{1}, Article \bibinfo{articleno}{39} (\bibinfo{date}{mar} \bibinfo{year}{2021}), \bibinfo{numpages}{27}~pages.
\newblock
\urldef\tempurl%
\url{https://doi.org/10.1145/3448098}
\showDOI{\tempurl}


\bibitem[Warden and Situnayake(2019)]%
        {warden2019tinyml}
\bibfield{author}{\bibinfo{person}{Pete Warden} {and} \bibinfo{person}{Daniel Situnayake}.} \bibinfo{year}{2019}\natexlab{}.
\newblock \bibinfo{booktitle}{\emph{Tinyml: Machine learning with tensorflow lite on arduino and ultra-low-power microcontrollers}}.
\newblock \bibinfo{publisher}{O'Reilly Media}.
\newblock


\bibitem[Xu et~al\mbox{.}(2020)]%
        {xu2020earbuddy}
\bibfield{author}{\bibinfo{person}{Xuhai Xu}, \bibinfo{person}{Haitian Shi}, \bibinfo{person}{Xin Yi}, \bibinfo{person}{WenJia Liu}, \bibinfo{person}{Yukang Yan}, \bibinfo{person}{Yuanchun Shi}, \bibinfo{person}{Alex Mariakakis}, \bibinfo{person}{Jennifer Mankoff}, {and} \bibinfo{person}{Anind~K. Dey}.} \bibinfo{year}{2020}\natexlab{}.
\newblock \showarticletitle{EarBuddy: Enabling On-Face Interaction via Wireless Earbuds}. In \bibinfo{booktitle}{\emph{Proceedings of the 2020 CHI Conference on Human Factors in Computing Systems}} (Honolulu, HI, USA) \emph{(\bibinfo{series}{CHI '20})}. \bibinfo{publisher}{Association for Computing Machinery}, \bibinfo{address}{New York, NY, USA}, \bibinfo{pages}{1–14}.
\newblock
\showISBNx{9781450367080}
\urldef\tempurl%
\url{https://doi.org/10.1145/3313831.3376836}
\showDOI{\tempurl}


\bibitem[Zhao et~al\mbox{.}(2007)]%
        {earpod07shengdong}
\bibfield{author}{\bibinfo{person}{Shengdong Zhao}, \bibinfo{person}{Pierre Dragicevic}, \bibinfo{person}{Mark Chignell}, \bibinfo{person}{Ravin Balakrishnan}, {and} \bibinfo{person}{Patrick Baudisch}.} \bibinfo{year}{2007}\natexlab{}.
\newblock \showarticletitle{Earpod: eyes-free menu selection using touch input and reactive audio feedback}. In \bibinfo{booktitle}{\emph{Proceedings of the SIGCHI Conference on Human Factors in Computing Systems}} (, San Jose, California, USA,) \emph{(\bibinfo{series}{CHI '07})}. \bibinfo{publisher}{Association for Computing Machinery}, \bibinfo{address}{New York, NY, USA}, \bibinfo{pages}{1395–1404}.
\newblock
\showISBNx{9781595935939}
\urldef\tempurl%
\url{https://doi.org/10.1145/1240624.1240836}
\showDOI{\tempurl}


\end{thebibliography}











\end{document}